\documentclass[preprint2]{aastex63}
\usepackage{xspace}
\usepackage{amsmath}
\newcommand{\edot}{\ensuremath{L_{\text{sd}}}\xspace}

\newcommand{\rsh}{\ensuremath{R_{\text{sh}}}\xspace}

\newcommand{\nh}{$N_{\rm H}$\xspace}

\newcommand{\msun}{\ensuremath{M_{\odot}}\xspace}

\newcommand{\src}{HESS~J0632$+$057\xspace}
\newcommand{\srca}{J0632\xspace}
\newcommand{\nustar}{\textit{NuSTAR}\xspace}
\newcommand{\swift}{{\it Swift-}XRT\xspace}

\def\asec{\ifmmode^{\prime\prime}\else$^{\prime\prime}$\fi}

\def\simgt{\lower.5ex\hbox{$\; \buildrel > \over \sim \;$}}
\def\simlt{\lower.5ex\hbox{$\; \buildrel < \over \sim \;$}}

\graphicspath{{./}{figures/}}

\shorttitle{\src}
\shortauthors{Tokayer et al.}

\begin{document}

\title{Multi-Wavelength Observation Campaign of the TeV Gamma-Ray Binary HESS J0632+057 with \nustar, VERITAS, MDM, and \emph{Swift}}

% \title{Contemporaneous Multi-Wavelength Campaign to Study HESS~J0632$+$057’s Distinctive Light Curve}

\author{Y.~M.~Tokayer}\affiliation{Columbia Astrophysics Laboratory, Columbia University, New York, NY, USA}
\author{H.~An}\affiliation{Department of Astronomy and Space Science, Chungbuk National University, Cheongju, 28644, Republic of Korea}
\author{J.~P.~Halpern}\affiliation{Columbia Astrophysics Laboratory, Columbia University, New York, NY, USA}
\author{J.~Kim}\affiliation{Department of Astronomy and Space Science, Chungbuk National University, Cheongju, 28644, Republic of Korea}
\author{K.~Mori}\affiliation{Columbia Astrophysics Laboratory, Columbia University, New York, NY, USA}
\author{C.~J.~Hailey}\affiliation{Columbia Astrophysics Laboratory, Columbia University, New York, NY, USA}
\collaboration{6}{(NuSTAR Collaboration, MDM)}

\author{C.~B.~Adams}\affiliation{Department of Physics and Astronomy, Barnard College, Columbia University, NY 10027, USA}
\author{W.~Benbow}\affiliation{Center for Astrophysics $|$ Harvard \& Smithsonian, Cambridge, MA 02138, USA}
\author{A.~Brill}\affiliation{Physics Department, Columbia University, New York, NY 10027, USA}
\author{J.~H.~Buckley}\affiliation{Department of Physics, Washington University, St. Louis, MO 63130, USA}
\author{M.~Capasso}\affiliation{Department of Physics and Astronomy, Barnard College, Columbia University, NY 10027, USA}
\author{M.~Errando}\affiliation{Department of Physics, Washington University, St. Louis, MO 63130, USA}
\author{A.~Falcone}\affiliation{Department of Astronomy and Astrophysics, 525 Davey Lab, Pennsylvania State University, University Park, PA 16802, USA}
\author{K.~A~Farrell}\affiliation{School of Physics, University College Dublin, Belfield, Dublin 4, Ireland}
\author{G.~M~Foote}\affiliation{Department of Physics and Astronomy and the Bartol Research Institute, University of Delaware, Newark, DE 19716, USA}
\author{L.~Fortson}\affiliation{School of Physics and Astronomy, University of Minnesota, Minneapolis, MN 55455, USA}
\author{A.~Furniss}\affiliation{Department of Physics, California State University - East Bay, Hayward, CA 94542, USA}
\author{A.~Gent}\affiliation{School of Physics and Center for Relativistic Astrophysics, Georgia Institute of Technology, 837 State Street NW, Atlanta, GA 30332-0430}
\author{C.~Giuri}\affiliation{DESY, Platanenallee 6, 15738 Zeuthen, Germany}
\author{D.~Hanna}\affiliation{Physics Department, McGill University, Montreal, QC H3A 2T8, Canada}
\author{T.~Hassan}\affiliation{DESY, Platanenallee 6, 15738 Zeuthen, Germany}
\author{O.~Hervet}\affiliation{Santa Cruz Institute for Particle Physics and Department of Physics, University of California, Santa Cruz, CA 95064, USA}
\author{J.~Holder}\affiliation{Department of Physics and Astronomy and the Bartol Research Institute, University of Delaware, Newark, DE 19716, USA}
\author{B.~Hona}\affiliation{Department of Physics and Astronomy, University of Utah, Salt Lake City, UT 84112, USA}
\author{T.~B.~Humensky}\affiliation{Physics Department, Columbia University, New York, NY 10027, USA}
\author{W.~Jin}\affiliation{Department of Physics and Astronomy, University of Alabama, Tuscaloosa, AL 35487, USA}
\author{P.~Kaaret}\affiliation{Department of Physics and Astronomy, University of Iowa, Van Allen Hall, Iowa City, IA 52242, USA}
\author{M.~Kertzman}\affiliation{Department of Physics and Astronomy, DePauw University, Greencastle, IN 46135-0037, USA}
\author{D.~Kieda}\affiliation{Department of Physics and Astronomy, University of Utah, Salt Lake City, UT 84112, USA}
\author{M.~J.~Lang}\affiliation{School of Physics, National University of Ireland Galway, University Road, Galway, Ireland}
\author{G.~Maier}\affiliation{DESY, Platanenallee 6, 15738 Zeuthen, Germany}
\author{C.~E~McGrath}\affiliation{School of Physics, University College Dublin, Belfield, Dublin 4, Ireland}
\author{P.~Moriarty}\affiliation{School of Physics, National University of Ireland Galway, University Road, Galway, Ireland}
\author{R.~Mukherjee}\affiliation{Department of Physics and Astronomy, Barnard College, Columbia University, NY 10027, USA}
\author{M.~Nievas-Rosillo}\affiliation{DESY, Platanenallee 6, 15738 Zeuthen, Germany}
\author{S.~O'Brien}\affiliation{Physics Department, McGill University, Montreal, QC H3A 2T8, Canada}
\author{R.~A.~Ong}\affiliation{Department of Physics and Astronomy, University of California, Los Angeles, CA 90095, USA}
\author{A.~N.~Otte}\affiliation{School of Physics and Center for Relativistic Astrophysics, Georgia Institute of Technology, 837 State Street NW, Atlanta, GA 30332-0430}
\author{N.~Park}\affiliation{Department of Physics, Engineering Physics \& Astronomy, Queen`s University, Kingston Ontario, Canada}
\author{S.~Patel}\affiliation{Department of Physics and Astronomy, University of Iowa, Van Allen Hall, Iowa City, IA 52242, USA}
\author{K.~Pfrang}\affiliation{DESY, Platanenallee 6, 15738 Zeuthen, Germany}
\author{M.~Pohl}\affiliation{Institute of Physics and Astronomy, University of Potsdam, 14476 Potsdam-Golm, Germany and DESY, Platanenallee 6, 15738 Zeuthen, Germany}
\author{R.~R.~Prado}\affiliation{DESY, Platanenallee 6, 15738 Zeuthen, Germany}
\author{E.~Pueschel}\affiliation{DESY, Platanenallee 6, 15738 Zeuthen, Germany}
\author{J.~Quinn}\affiliation{School of Physics, University College Dublin, Belfield, Dublin 4, Ireland}
\author{K.~Ragan}\affiliation{Physics Department, McGill University, Montreal, QC H3A 2T8, Canada}
\author{P.~T.~Reynolds}\affiliation{Department of Physical Sciences, Munster Technological University, Bishopstown, Cork, T12 P928, Ireland}
\author{D.~Ribeiro}\affiliation{Physics Department, Columbia University, New York, NY 10027, USA}
\author{E.~Roache}\affiliation{Center for Astrophysics $|$ Harvard \& Smithsonian, Cambridge, MA 02138, USA}
\author{J.~L.~Ryan}\affiliation{Department of Physics and Astronomy, University of California, Los Angeles, CA 90095, USA}
\author{M.~Santander}\affiliation{Department of Physics and Astronomy, University of Alabama, Tuscaloosa, AL 35487, USA}
\author{S.~Schlenstedt}\affiliation{CTAO, Saupfercheckweg 1, 69117 Heidelberg, Germany}
\author{G.~H.~Sembroski}\affiliation{Department of Physics and Astronomy, Purdue University, West Lafayette, IN 47907, USA}
\author{A.~Weinstein}\affiliation{Department of Physics and Astronomy, Iowa State University, Ames, IA 50011, USA}
\author{D.~A.~Williams}\affiliation{Santa Cruz Institute for Particle Physics and Department of Physics, University of California, Santa Cruz, CA 95064, USA}
\author{T.~J~Williamson}\affiliation{Department of Physics and Astronomy and the Bartol Research Institute, University of Delaware, Newark, DE 19716, USA}
\collaboration{55}{(VERITAS Collaboration)}

\correspondingauthor{Y.~M. Tokayer}
\email{y.tokayer@columbia.edu}

\correspondingauthor{R.~R. Prado}
\email{raul.prado@desy.de}

\begin{abstract}

% \src is a TeV gamma-ray binary that stands out for its distinctive, double-peaked, high energy light curve.  Despite extensive observations in X-rays and gamma-rays, a quantitative explanation of this orbitally modulated flux has remained elusive, as has a unique orbital solution. We reanalyze over ten years of archived \swift data and present a novel geometrically-motivated light curve model that accounts for beaming effects, B-field modulation at the intra-binary shock, and pulsar passage through the Be star disk. The model constrains the system's orbit, and robustly determines the phases of the periastron, disk crossings, and inferior conjunction. orbital parameters.  We also present the results of quasi-simultaneous observations of \src in the optical, X-ray, and gamma-ray bands.  The spectral analysis and updated SED fit are consistent with our light curve model.

%We report a multi-wavelength observation campaign of TeV gamma-ray binary HESS J0632+057 with NuSTAR, VERITAS and MDM observatory. 
\src\ belongs to a rare subclass of binary  systems which emits gamma-rays above 100 GeV. It stands out for its distinctive high-energy light curve, which features a sharp ``primary'' peak and broader ``secondary'' peak.
% The high-energy emission originates from particle acceleration at the shock formed by the wind-wind collision of the Be star and pulsar companion.
We present the results of contemporaneous observations by \nustar and VERITAS during the secondary peak between Dec.\ 2019 and Feb.\ 2020, when the orbital phase ($\phi$) is between 0.55 and 0.75.
\nustar\ detected X-ray spectral evolution, while VERITAS detected TeV emission.
We fit a leptonic wind-collision model to the multi-wavelength spectra data obtained over the four \nustar\ and VERITAS observations, constraining the pulsar spin-down luminosity and the magnetization parameter at the shock.
Despite long-term monitoring of the source from Oct.\ 2019 to Mar.\ 2020, the MDM observatory did not detect significant variation in H$\alpha$ and H$\beta$ line equivalent widths, an expected signature of Be-disk interaction with the pulsar.
Furthermore, fitting folded \swift light curve data with an intra-binary shock model constrained the orbital parameters, suggesting two orbital phases (at $\phi_D = 0.13$ and 0.37) where the pulsar crosses the Be-disk, as well as phases for the periastron ($\phi_0 = 0.30$) and inferior conjunction ($\phi_{\text{IFC}} = 0.75$).
The broad-band X-ray spectra with \swift and \nustar\ allowed us to measure a higher neutral hydrogen column density at one of the predicted disk-passing phases.

\end{abstract}

\keywords{gamma rays: general -- stars: individual (MWC 148) -- X-rays: binaries -- X-rays: individual (\src) }

%%%%%%%%%%%%%%%%%%%%%%%%%%%%%%%%%%%%%%%%%%%%%%%%%%%%%
\section{Introduction} % Yarone
\label{sec:intro}

Over the past two decades, ground-based gamma-ray telescopes, together with X-ray telescopes, have uncovered a rare subclass of binary systems detected at energies $>$100 GeV, eight of which have been unambiguously discovered to date \citep{Chernyakova-2020}.
Each of these so-called TeV gamma-ray binaries (TGBs) consists of an O or B main sequence star and a compact object companion, with orbital periods ranging from 3.9 days in the case of LS 5039 \citep{Casares-2005} to $\sim$50 years in the case of PSR J2032$+$4127 \citep{Ho-2017}.

%Two types of models have been invoked to explain the high-energy non-thermal emission of these binary systems, depending on the nature of the system's compact object: the ``microquasar scenario'' and the ``pulsar scenario.''
%In the microquasar scenario, it is speculated that, in an analogy to AGNs, material from the star accretes onto the compact object (a black hole), with some particles accelerated along relativistic jets.
%In the case of a pulsar, strong winds from a young pulsar prevent accretion, and a shock boundary forms with the circumstellar material of the main sequence star, accelerating particles into the TeV range.
%Radio pulsars have been directly observed in two TGBs (PSR  B1259$-$63  and  PSR  J2032$+$4127).
%While the nature of the compact object in the remaining systems is still unknown, indirect evidence has favored the pulsar scenario \citep{Dubus-2013}.

\src (henceforth ``\srca'') was first detected as an unidentified point-like source during H.E.S.S. observations of the Monoceros region \citep{Aharonian-2007}, and long term ($\sim100$ days) X-ray and gamma-ray variability provided important evidence of its binary nature \citep{Acciari-2009}. A follow-up X-ray monitoring of \srca\ detected an orbital period of $321\pm5$ days \citep{Bongiorno-2011}.
Its optical counterpart is the Be star MWC 148, which, through optical spectroscopy, was estimated to be at a distance of 1.1--1.7 kpc \citep{Aragona-2010}.
As has been done in previous X-ray studies \citep[e.g.,][]{Aliu-2014, Archer-2020}, we adopt a value of 1.4 kpc in this paper.
While \srca is luminous in the TeV and X-ray bands, it is uncommonly faint in the GeV band \citep{Li-2017}.

One model that is invoked to explain the high energy non-thermal emission of TGBs is the ``pulsar scenario,'' in which strong winds from a young pulsar prevent accretion, and a shock boundary forms with the circumstellar material of the Be star, accelerating particles into the TeV range.
Although no pulsation has yet been detected, \cite{Moritani-2018} argued for the pulsar scenario in \srca, since the mass function derived from H$\alpha$ velocities of the Be star indicates that the compact object mass is $<2.5\;\msun$, which is consistent with a neutron star.
% Furthermore, \srca has a distinctive orbital light curve that resembles the TGB 1FGL J1018.6$-$5856 (see next paragraph), which X-ray studies suggest hosts a pulsar \citep{An-2015, Waisberg-2015}. Finally, 
Furthermore, a previous study \citep{Archer-2020} showed that the spectral energy distribution (SED) of combined X-ray and gamma-ray emission from \srca is consistent with the pulsar scenario. These indications, together with the assumption of similarity among all objects of this class, two of which have known pulsars, lead us to adopt the pulsar scenario for \srca.

\srca features a distinctive double-peaked orbital light curve in both the X-ray and gamma-ray bands, with significant gamma-ray flux variation (which is characteristic of TGBs) \citep{Acciari-2009}.
The light curve consists of a tall, narrow peak (the ``primary peak'') followed by a sharp drop-off and a smaller broad peak (the ``secondary peak'').
% Exactly where in the orbit these peaks lie depends on which orbital period is adopted.
\cite{Moritani-2015} explained the high-energy light curve in terms of a ``flip-flop'' scenario (originally proposed by \cite{Torres-2012} for the TGB LS I+61 303), in which when the pulsar is close to periastron (at phase $\phi$ $\sim$ 0, according to the orbital solution of \cite{Casares-2012}), the strong gas pressure quenches the pulsar wind, thus suppressing the high-energy emission.
The second minimum corresponds to apastron, where the magnetic field at the shock boundary is low, as is the field of soft photons from the Be star.
\cite{Malyshev-2019} instead adopt an ``inclined disk'' model (originally proposed by \cite{Chernyakova-2015} for the TGB PSR B1259$-$63), in which the Be star's disk is inclined relative to the orbital plane of the neutron star, and the two peaks of the light curve are explained by passage through the disk, where the higher density leads to enhanced acceleration at the shock boundary.
The primary peak occurs at a transit closer to periastron (which, according to their orbital solution is at $\phi$ $\sim$ 0.4), where the disk is concentrated and narrow, and the pulsar is faster, while the secondary peak occurs farther out, where the disk is more diffuse and splayed, and the pulsar velocity is decreased.
This model asserts that our line of sight to the system is oriented ``edge on'' relative to the circumstellar disk.
Therefore, at peak light curve phases, it predicts signatures of disk disruption in both the optical (as a modulation in the H$\alpha$ line) and X-ray (as an increase in the hydrogen column density) spectra.
%It therefore predicts indications of disk disruption in the optical spectrum, as well as in the X-ray spectrum as an increase in the hydrogen column density at the peak light curve phases.
% We test and discuss these predictions in Sec.~\ref{subsec:mdm_analysis} and~\ref{sec:lc}.

While the flip-flop scenario and inclined disk model attempt to explain the light curve's overall dips and peaks, respectively, neither has been used to quantitatively model the X-ray flux of \srca over the orbital period, nor have they been invoked to explain the light curve's more detailed features, such as the small excess before the primary peak and the sharp drop off after the primary peak.
Meanwhile, 1FGL J1018.6-5856, another TGB, exhibits a very similar double-peak structure---one narrow, one broad---in its X-ray light curve \citep{An-2015}. \cite{An-2017} quantitatively explained this using a geometrically-motivated emission model.
%This leads us to explore if a similar model, modified to account for \srca's unique properties, can be used to explain \srca's light curve (Sec.~\ref{subsec:lc_model} and~\ref{subsec:lc_results}).
Extensive X-ray observations covering the entire orbit of \srca present an opportunity to apply a similar model.

Unlike for other TGBs, a unique orbital solution for \srca has not yet been established.
Two distinct scenarios have been derived by \cite{Casares-2012}, using optical data, and \cite{Moritani-2018}, using both optical and soft X-ray data.
Both have been found to be consistent with high-energy studies \citep[e.g.,][]{Malyshev-2019, Archer-2020}.
As for the orbital period, \cite{Maier-2019} present the most precise orbital period based on \swift data, $317.3 \pm 0.7$ days, which they found to be consistent with gamma-ray light curves derived by H.E.S.S., MAGIC, and VERITAS observations.
%An elaboration of the period determination can be found in a forthcoming paper by those collaborations.

A shortcoming of previous spectral studies of \srca in X-rays is that, despite long-term monitoring by \swift, observations below 10 keV suffer from degeneracy between the hydrogen column density (\nh) and the photon index, both of which may vary with orbital phase.
% Our previous study~\citep{Archer-2020} aimed to remedy this problem with simultaneous \nustar and VERITAS observations in Nov.\ and Dec.\ of 2017, during the primary peak of the high-energy light curve.
Our previous study~\citep{Archer-2020} aimed to remedy this problem with \nustar observations in Nov.\ and Dec.\ of 2017, during the primary peak of the high-energy light curve.
Because \nustar features sensitivity in hard X-rays, its spectra have little to no dependency on \nh ($\sim5\times10^{21}$ cm$^{-2}$ for \srca).
The observations yielded precise measurements of the photon index in both the X-ray and gamma-ray bands.
The spectra of both bands were harder in Dec.~than in Nov., suggesting emission from a single electron population.
That study constrained the pulsar spin-down luminosity ($\edot$) and the magnetization parameter ($\sigma$) at the shock, based on a joint SED fitting of the X-ray and gamma-ray observations, and showed that broadband spectroscopy with \nustar, combined with TeV observations, is effective for determining fundamental system parameters.
The magnetization parameter is the ratio of the Poynting flux to the matter flux of the pulsar outflow, defined by $\sigma = F_p/F_m = B^2/4\pi\Gamma\rho c^2$, where $B$ is the magnetic field strength and $\rho$ is the matter density, both in the observer frame.
Constraints on $\sigma$ are crucial for understanding how energy from a pulsar is transferred to its surroundings.
While the pulsar wind is believed to be dominated by the Poynting flux near the light cylinder ($\sigma\gg1$), and observations of the Crab Nebula constrain $\sigma\ll1$ at the pulsar wind termination shock, TGBs present an opportunity to measure $\sigma$ at intermediate distances \citep{Kirk-2009}.
However, while the pulsar/stellar wind parameters were constrained, they were not uniquely determined, largely because the 2017 observations covered only a small fraction of the orbit around the primary peak.

This paper presents simultaneous observations of \srca across multiple energy bands.
They add to our previous data set by observing \srca during the secondary peak of the light curve in the hard X-ray (\nustar), and gamma-ray (VERITAS) bands, with the addition of quasi-simultaneous optical observations by the MDM observatory. %in order to look for indications of disruptions of the Be star disk by the compact object.
The second observations by \nustar and VERITAS were originally intended to be triggered by H$\alpha$ modulation (which would indicate disk-disruption during the secondary peak), but, when no modulation was detected, the observations were carried out while \srca was still observable to VERITAS.
The present study also utilized the collection of 273 archived observations of \srca in soft X-rays by the \swift telescope, which are dated between 2009-01-26 and 2020-02-23.

Using the results of our analyses, we present an explanation of \srca's double-peaked X-ray light curve that takes into account its orbital geometry, the orbital modulation of system parameters such as the $B$-field strength at the shock, and the interaction between the pulsar and the circumstellar disk.
% (but with disk-passages occurring at different phases than those predicted by~\cite{Malyshev-2019}).
We also propose a new orbital solution consistent with our light curve and X-ray data analysis.
%in Sec.~\ref{subsec:lc_results}.
%This explanation poses new constraints on \srca's orbital solution.
Finally, we present an updated broadband high-energy SED fit, which gives constraints on $\edot$ of \srca's pulsar and $\sigma$ at the location of the shock, using parameters from our proposed orbital solution.
We also include the orbital solutions of~\cite{Casares-2012} and~\cite{Moritani-2018} in the SED fit, and discuss points of divergence between them.
% This model explains the non-thermal emission as the result of electrons from the pulsar wind being accelerated at the wind-wind shock boundary.
% The hard X-ray and TeV gamma-ray spectra are assumed to be produced through synchrotron radiation and inverse Compton scattering of stellar photons, respectively.
%This model does not include a circumstellar disk component, which is justified for the orbital phases of the \nustar and VERITAS observation by the orbital solution we propose. % This may no longer be true...

Observations and data analysis are described in Sec.~\ref{sec:observations} and~\ref{sec:data_analysis}.
In Sec.~\ref{sec:lc} we describe the model used to fit the X-ray light curve and present and discuss its results.
Finally, we present our joint-SED fit in Sec.~\ref{sec:sed}, and our conclusions in Sec.~\ref{sec:conclusions}.

%%%%%%%%%%%%%%%%%%%%%%%%%%%%%%%%%%%%%%%%%%%%%%%%%%%%%
\section{Observations}
\label{sec:observations}

\subsection{\swift} % Yarone

The Neil Gehrels \emph{Swift} Observatory's X-ray Telescope (\swift) is a low-Earth orbiting telescope that is sensitive to soft X-rays (0.2--10 keV) \citep{Burrows-2005}.
% It has an angular resolution of $18\asec$ (HPD), and energy resolution of 140 eV at 5.9 keV, and a timing resolution $\leq$~10~ms \citep{Burrows-2005}.
There are 273 publicly available observations of \srca (ObsID prefixes 00031329xxx, 00088078xxx, 00088643xxx, 00088913, and 00090417xxx) that took place between 2009-01-26 and 2020-02-23, and whose exposures range from $\sim$50 sec to $\sim$7.5 ks (a typical exposure is 4--5 ks).
These observations include those that are simultaneous with the four \nustar observations of \srca.

\subsection{\nustar} % Yarone
\label{subsec:nustar_obs}

The \nustar observatory consists of two co-aligned grazing-incidence X-ray telescopes with focal plane modules FPMA and FPMB.
It has an imaging resolution of 18\asec\ FWHM and 58\asec\ HPD over an energy band of 3--79 keV and a characteristic 400 eV FWHM spectral resolution at 10 keV \citep{Harrison-2013}.
The absolute and relative timing precision of \nustar, after correcting for on-board clock drift, are 3 msec \citep{Madsen-2015} and 10 $\mu$sec \citep{Bachetti-2015}, respectively.
\nustar's broadband capabilities allow it to measure spectral properties such as photon index with relatively high precision, with minimal dependence on interstellar medium absorption.

The \nustar observations used in this study took place on 2017-11-22 (ObsID 30362001002; 49.7~ks), 2017-12-14 (ObsID: 30362001004; 49.6~ks), 2019-12-22 (ObsID: 30502017002; 43.0~ks), and 2020-02-22 (ObsID: 30502017004; 53.1~ks).
The first pair of observations (henceforth ``Nu1a'' and ``Nu1b'') provided the X-ray data in our first study~\citep{Archer-2020}, and corresponded to the rise of the primary peak in \src's double-peaked light curve.
The second pair of observations (``Nu2a'' and ``Nu2b'') were designed to take place approximately halfway across the orbit from the first two observations. See Fig.~\ref{fig:swift_lc}.

\subsection{VERITAS} %Gernot
\label{subsec:veritas_obs}

The VERITAS observatory is an array of four 12m-diameter ground-based imaging atmospheric Cherenkov telescopes located at the Fred Lawrence Whipple Observatory in southern Arizona (1300~m above sea level, N$31^\circ 40\arcmin 30\arcsec$, W$110^\circ 57\arcmin 08\arcsec$; \citet{Holder-2006}).
VERITAS detects Cherenkov light emitted from extensive air showers, which are initiated by the interaction of high-energy gamma-ray showers with the atmosphere.
The effective area for gamma rays in the TeV energy range is about $10^5$ m$^2$. 
VERITAS can detect a source with a flux level of 1\% of the steady flux from the Crab Nebula in less than 25 hr. 
The angular resolution, defined as the 68\% containment radius of the instrument at 1 TeV, is better than $0.1^{\circ}$ \citep{Park-2015}.

% 201912 414.33 min, 2019, Dec 20 to 2020, Jan 3
% 202001 468.82,om, 2020, Jan 19 to 2020, Jan 30
% 202002 500.22min, 2020, Feb 18 to 2020, Feb 28
VERITAS observed \srca for 6.9 hours between 2019, Dec 20 and 2020, Jan 3 (``Ve2a''); for 7.8 hours between 2020, Jan 19 and 30 (``Ve2b''); and for 8.3 hours between 2020, Feb 18 and 28 (``Ve2c'').
The elevation range for the observations varied between 49 and 62$^{\circ}$, resulting in an energy threshold varying between 200 and 350 GeV.
Two additional VERITAS observations of \srca, Ve1a and Ve1b, took place in Nov-Dec of 2017, as previously reported in \cite{Archer-2020}.
 
\subsection{MDM Observatory Spectroscopy} %Jules
\label{subsec:mdm_obs}

Optical spectra of MWC 148 were obtained in queue mode on the 2.4~m Hiltner telescope of the MDM Observatory on Kitt Peak, Arizona.
The Ohio State Multi-Object Spectrograph \citep{mar11} was used with a volume-phase holographic grism and a single $1.2\asec$ wide slit, which provided a dispersion of 0.72~\AA\ pixel$^{-1}$ and a resolution of $\approx2.5$~\AA\ over the wavelength range 3970--6880~\AA.
On each of 12 nights between 2019 October and 2020 March ($0.3 < \phi < 0.8$), three spectra with exposure times of 30--60 seconds were recorded.
Wavelength calibration was carried out using arc lamp comparison spectra taken at the beginning or end of each night. The orbital phases of the MDM observations are indicated in Fig.~\ref{fig:swift_lc}.

%  A log of the observations and basic results are given in Table~\ref{tab:optlog}.

%%%%%%%%%%%%%%%%%%%%%%%%%%%%%%%%%%%%%%%%%%%%%%%%%%%%%
\section{Data Analysis}
\label{sec:data_analysis}

\subsection{\swift Period Search and Light Curve} % Yarone
\label{subsec:swift_analysis}

\begin{figure}
\includegraphics[angle=0,width=85mm]{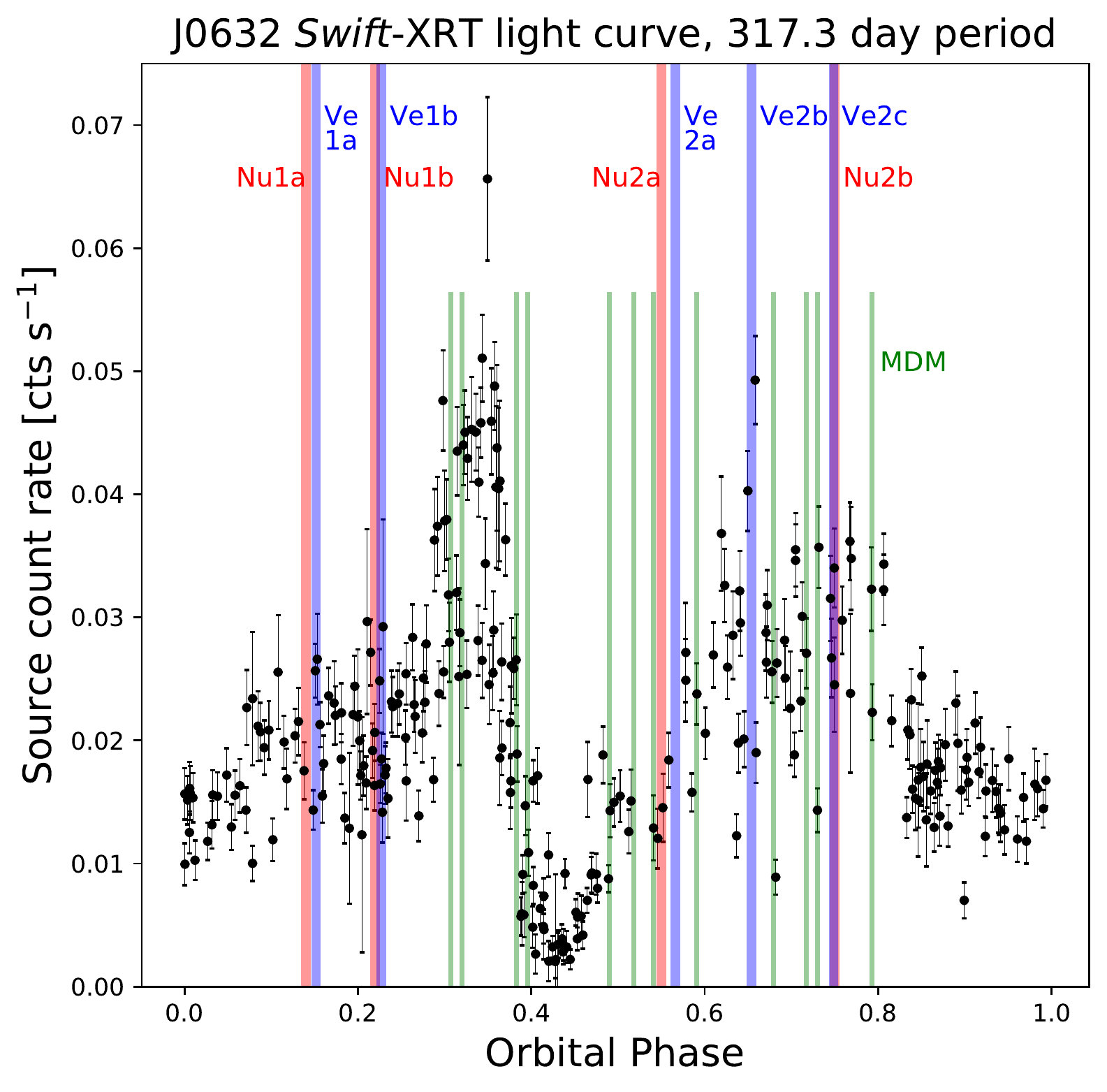}
\caption{\label{fig:swift_lc}
Swift light curve folded to an orbital period of 317.3 days, with $t_0$ = MJD 54857.0, the date of the first \swift observation.  Each data point represents one \swift observation.  1$\sigma$ errors are shown.  Vertical lines indicate the phases of the \nustar (red), VERITAS (blue), and MDM (green) observations.
swiftNu2, Ve2, and MDM observations were all during the same 2019-2020 orbit, while the Nu1 and Ve1 observations were of the same orbit in late 2017.}

\end{figure}

We adopt the orbital period of $317.3 \pm 0.7$ days presented by \citet{Maier-2019}, which was derived from \swift observations through 2019-01, and the methods of which are detailed in a forthcoming paper by that collaboration \citep{Maier-2021}.
To construct a folded light curve, all data (observations through 2019-02) were reprocessed with \texttt{swiftpipeline} v0.13.5.
Barycentric correction was applied using the HEASoft (v6.28) \texttt{barycorr} tool.
Event files were filtered by energy (0.2--10 keV) and $30\asec$ circular regions centered on $RA$ = 6:32:59.3 and $DEC$ = +5:48:01.4 (J2000) were extracted.
% To construct a folded light curve, source photons were counted from the same $30\asec$ circular regions used in the timing analysis.
Since the \swift angular resolution is 18\asec (half-power diameter, HPD), these regions reasonably contain all source counts.
Background counts were taken from concentric annular regions with $r_{\text{in}} = 50\asec$ and $r_{\text{out}} = 100\asec$.
Counts in the background regions were scaled by the ratio of the source area to background area, and subtracted from the source counts. 
We then divided by the total ``good'' exposure time to obtain a count rate for each observation. 
% 1$\sigma$ Poisson errors were calculated.
The folded light curve is shown in Fig.~\ref{fig:swift_lc}.

%%%%%%%%%%%%%%
%%%%%%%%%%%%

\subsection{\nustar} % Yarone
\label{subsec:nustar_analysis}

\subsubsection{\nustar Spectral Analysis}
The details of the spectral analysis for Nu1a and Nu1b can be found in \cite{Archer-2020}.
Data processing and analysis were performed using the HEASoft (v6.28) software package, including \texttt{NuSTARDAS} 06Jul17$\_$v1.8.0 and  and \nustar\ Calibration Database (CALDB) files dated 2019-12-19.
Source photons for Nu2a and Nu2b were extracted from circular regions centered around the brightest pixel position.
% Source photons for Nu2a and Nu2b were extracted from circular regions centered near $RA = 98.246^\circ$ and $DEC = +5.8018^\circ$ (regions were centered around the brightest pixel on each \nustar detector, which varied to be within the \nustar spatial resolution of the aforementioned coordinates).
Since \nustar observations were focused on the hard X-ray band, radii between $49\asec$ and $53\asec$ were chosen to maximize the signal-to-noise ratio of photons in the 10--30 keV range (background was found to be spatially uniform close to the source).
These regions yielded a net count of 3392 and 4259 photons for Nu2a and Nu2b, respectively (FPMA and FPMB combined), which translates to a count rate of 0.041/0.038 cts~s$^{-1}$ for Nu2a and 0.039/0.041 cts~s$^{-1}$ for Nu2b (FPMA/FPMB).
For all observations, background spectra were extracted from a rectangular source-free region on the same detector chip as the source.
Spectra, response matrix files  (\texttt{.rmf}) and ancillary response files (\texttt{.arf}) were all generated using the \texttt{nuproducts} command. 

\begin{figure}
\centering
\includegraphics[width=0.9\linewidth]{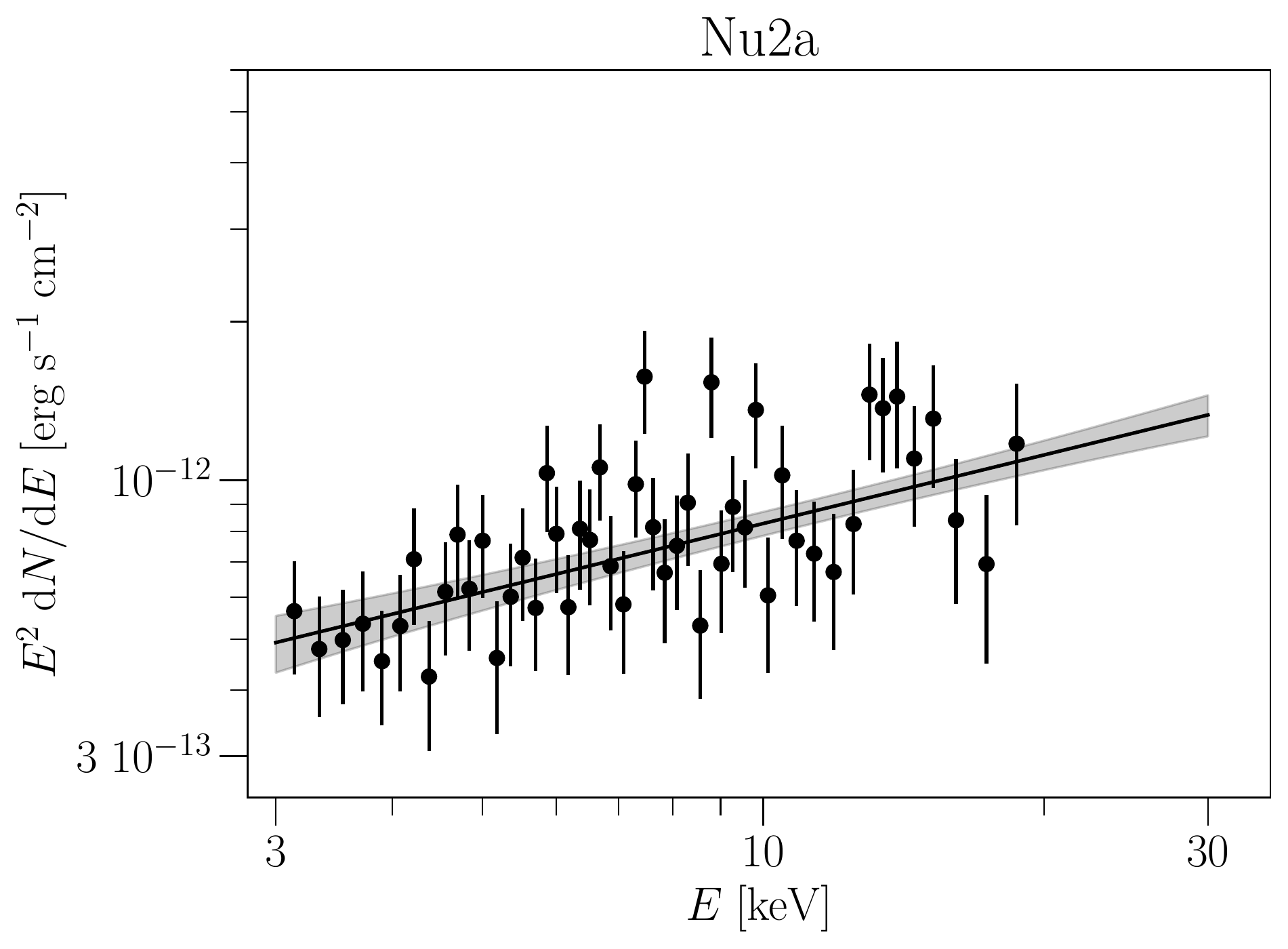}

\includegraphics[width=0.9\linewidth]{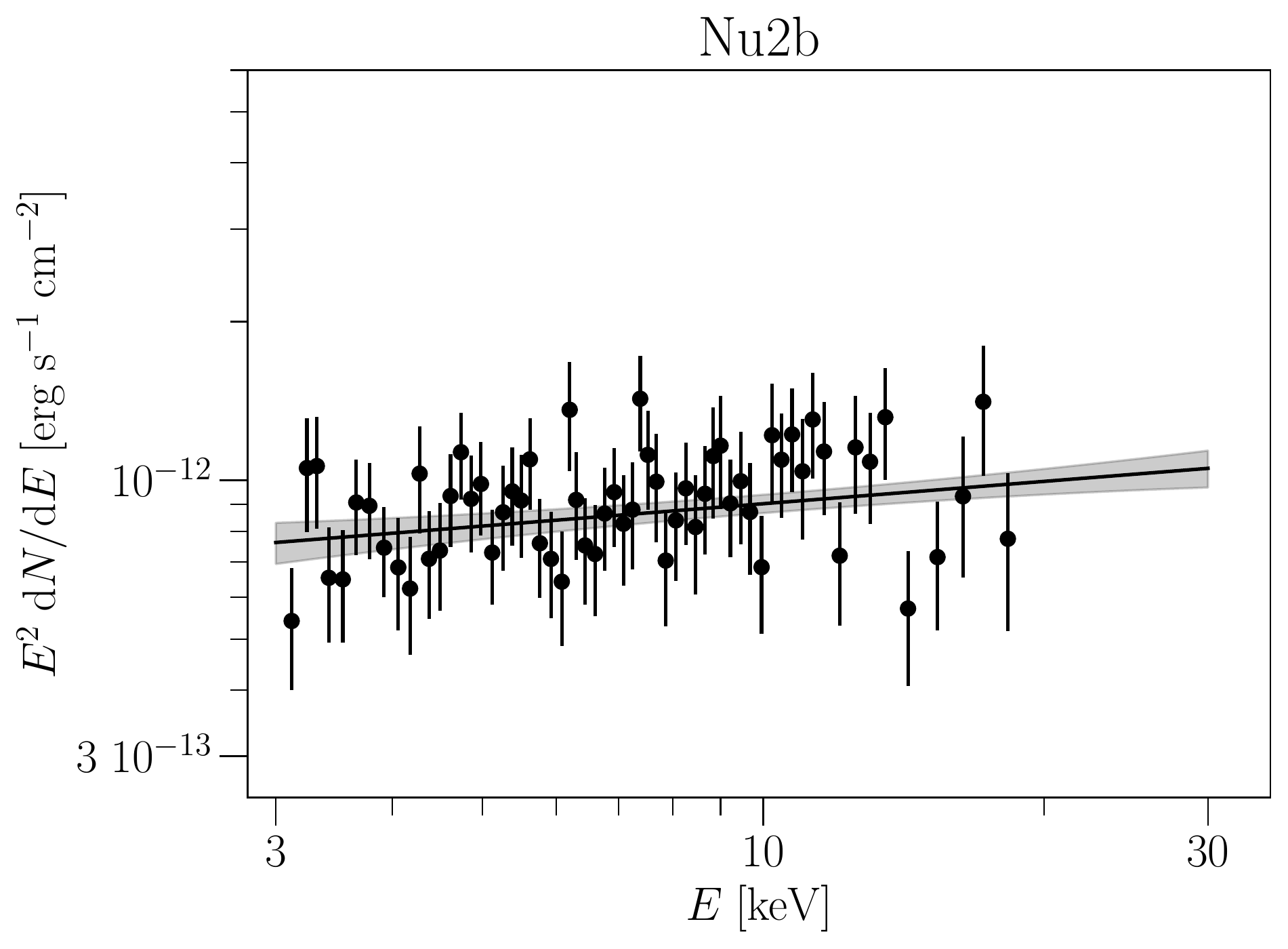}
\caption{\label{fig:data-sed-nustar}
SED derived from \nustar observations. The solid lines show the result of the single power law fit and the gray band its 1$\sigma$ confidence interval. Values of the integrated flux and spectral index can be found in Table~\ref{tab:nustar-spectral}.}

\end{figure}

Spectra were grouped to ensure a minimum of 20 counts per bin and fitted using \texttt{XSPEC} (v12.11.0) \citep{Arnaud-1996}. Spectra from FPMA and FPMB were jointly fit in the 3.0--20 keV range, above which the background was found to dominate.
\nustar spectra Nu2a and Nu2b were fit to a single power law model, the results of which are listed in Table \ref{tab:nustar-spectral}, together with the results of Nu1a and Nu1b.
We confirmed that interstellar medium (ISM) absorption is negligible in the \nustar band by also fitting the spectra to an absorbed power law model (\texttt{tbabs*powerlaw}), and observing no change in the best-fit spectral index ($\Gamma$) values, while best-fit \nh values spanned multiple orders of magnitude.
This is consistent with previously measured \nh values \citep[e.g.,][]{Moritani-2018}, which are not high enough to affect X-ray emission above 3 keV.
While spectral hardening was observed between Nu1a and Nu1b, the spectrum softened from Nu2a to Nu2b. The SEDs derived from Nu2a and Nu2b are shown in Fig.~\ref{fig:data-sed-nustar}.

\begin{deluxetable*}{c c c c c}
    \tablecolumns{5}
    \centering
    \tablecaption{\nustar spectral analysis results.  All statistical uncertainties are the 90\% confidence level using $\chi^2$ statistics.  Systematic errors are due to different flux normalizations between FPMA and FPMB.  Luminosities are derived from the flux values assuming a distance of 1.4 kpc. \label{tab:nustar-spectral}}
    
    \tablehead{
        \colhead{Observation} & \colhead{Photon Index, $\Gamma$} & \begin{tabular}{@{}c@{}}
            \colhead{3.0--20 keV Flux} \\
            \colhead{($10^{-12}$ erg cm$^{-2}$ s$^{-1}$)}
        \end{tabular} & \begin{tabular}{@{}c@{}}
            \colhead{Luminosity} \\
            \colhead{($10^{32}$ erg s$^{-1}$)}
        \end{tabular} & \colhead{red. $\chi^2$, d.o.f.}
        }
    \startdata
        Nu1a* & 1.77 $\pm$ 0.09 & $1.90$ $^{+0.08}_{-0.13}$ {\scriptsize (stat)} $\pm0.12$ {\scriptsize(sys)} & $4.46$ $^{+0.19}_{-0.30}$ {\scriptsize(stat)} $\pm0.28$ {\scriptsize(sys)}& 0.92, 105 \\
        Nu1b* & 1.56 $\pm$ 0.08 & $1.87$ $^{+0.09}_{-0.14}$ {\scriptsize (stat)} $\pm0.04$ {\scriptsize(sys)}& $4.39$ $^{+0.21}_{-0.33}$ {\scriptsize (stat)} $\pm0.09$ {\scriptsize(sys)}& 0.71, 102 \\\hline
        Nu2a & $1.57\pm0.10$ & $1.48$ $^{+0.09}_{-0.11}$ {\scriptsize (stat)} $\pm0.03$ {\scriptsize(sys)}& $3.47$ $^{+0.21}_{-0.26}$ {\scriptsize(stat)} $\pm0.07$ {\scriptsize(sys)} & 1.01, 95 \\
        Nu2b & $1.79\pm0.08$ & $1.69$ $^{+0.09}_{-0.11}$ {\scriptsize(stat)} $\pm0.01$ {\scriptsize(sys)}& $3.96$ $^{+0.21}_{-0.26}$ {\scriptsize(stat)} $\pm0.02$ {\scriptsize(sys)} & 0.92, 124
    \enddata
    \tablenotetext{*}{Values for Nu1a and Nu1b adopted from \cite{Archer-2020}.  Flux values and errors were recalculated for consistency with our data presentation.}
\end{deluxetable*}

%%%%%%%%%%%%%%
%%%%%%%%%%%%

\subsubsection{\nustar Timing Analysis}

To search for indications of accretion and/or signs of pulsation, we analyzed the power spectra of the \nustar observations.
Barycentric correction with the \nustar clock file was applied to the event files for both observations, and events were extracted from the same source regions used in our \nustar spectral analysis.
Event files were further filtered to the 3--20 keV energy range.
Light curves and power density spectra were generated using the \texttt{Stingray} Python library \citep{Stingray-2016}.
Given that the count rates for our observations were $\leq$ 0.041 cts/s for each module, the \nustar deadtime effect is negligible.
The filtered events from both modules were combined for each observation, and light curves were binned to a constant interval of $\Delta T = 0.016$~sec.
Thus we searched for frequencies up to 31.25 Hz (or periods down to 32 msec).

The resulting power density spectra were flat for both Nu2a and Nu2b (with variances of 0.0160 and 0.0126, respectively, in the Leahy normalized power).
For each of the searched frequencies, upper limits on sinusoidal modulation were calculated using \texttt{Stringray}’s adaptation of Eq. 10 from \cite{Vaughan-1994}, $a = 1.61\sqrt{\frac{P}{N_\gamma}}$, where $P$ is the upper limit on signal power derived from a noncentral chi-squared distribution, and $N_\gamma$ is the number of counts in the observation.
Amplitude upper limits were converted into pulse fractions by $\frac{2a}{1+a}$.
Pulse fraction upper limits on all searched frequencies were 0.164 and 0.163 (90\% conf.).  Thus, there was no sign of red noise or pulsations, which would suggest X-ray emission due to accretion or pulsed emission from the compact object.
%indicating that the X-ray emission is not due to accretion or pulsed emission from the compact object.
This is consistent with previous \nustar observations of \srca \citep{Archer-2020}.

\subsection{VERITAS Analysis} % Gernot
\label{subsec:veritas_analysis}

The VERITAS data processing pipeline consists of pixel and throughput calibration, image processing using second moment parameters, and stereo event reconstruction taking into account the information from different telescopes to determine photon energy and direction \citep{Acciari-2008,Maier-2017}. 
The applied background rejection method utilizes boosted decision trees \citep{Krause-2017}.
Pre-defined cuts optimised for point-like sources are applied to reject cosmic-ray background events.
The reflected-region method \citep{Fomin-1994} was used for the estimation of the remaining background after gamma-hadron separation cuts. The VERITAS data analysis is described in detail in \citet{Acciari-2008}; all results were cross-checked with an independent analysis chain \citep{Cogan-2007}.

\begin{figure}
\centering
\includegraphics[width=0.9\linewidth]{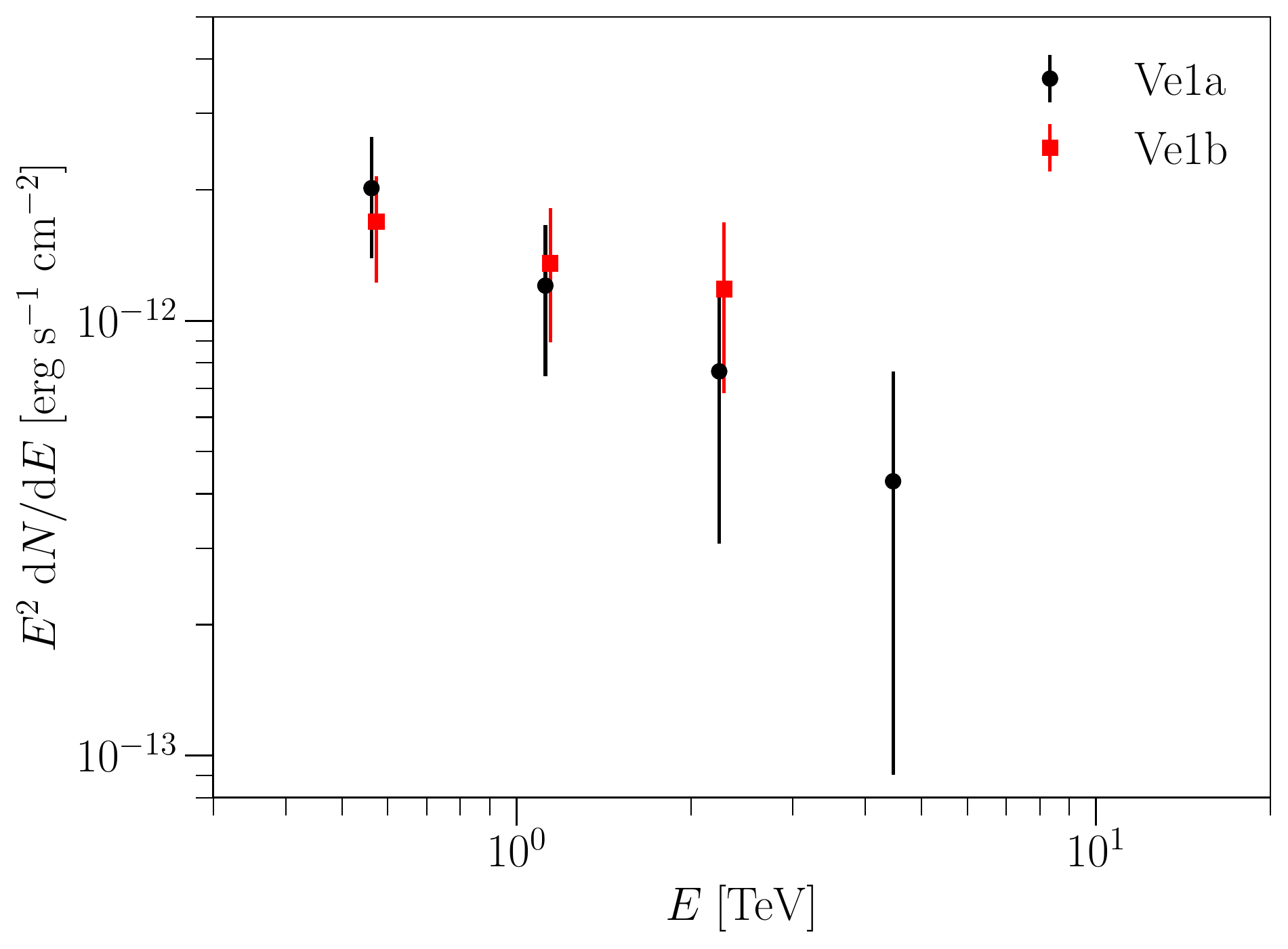}

\includegraphics[width=0.9\linewidth]{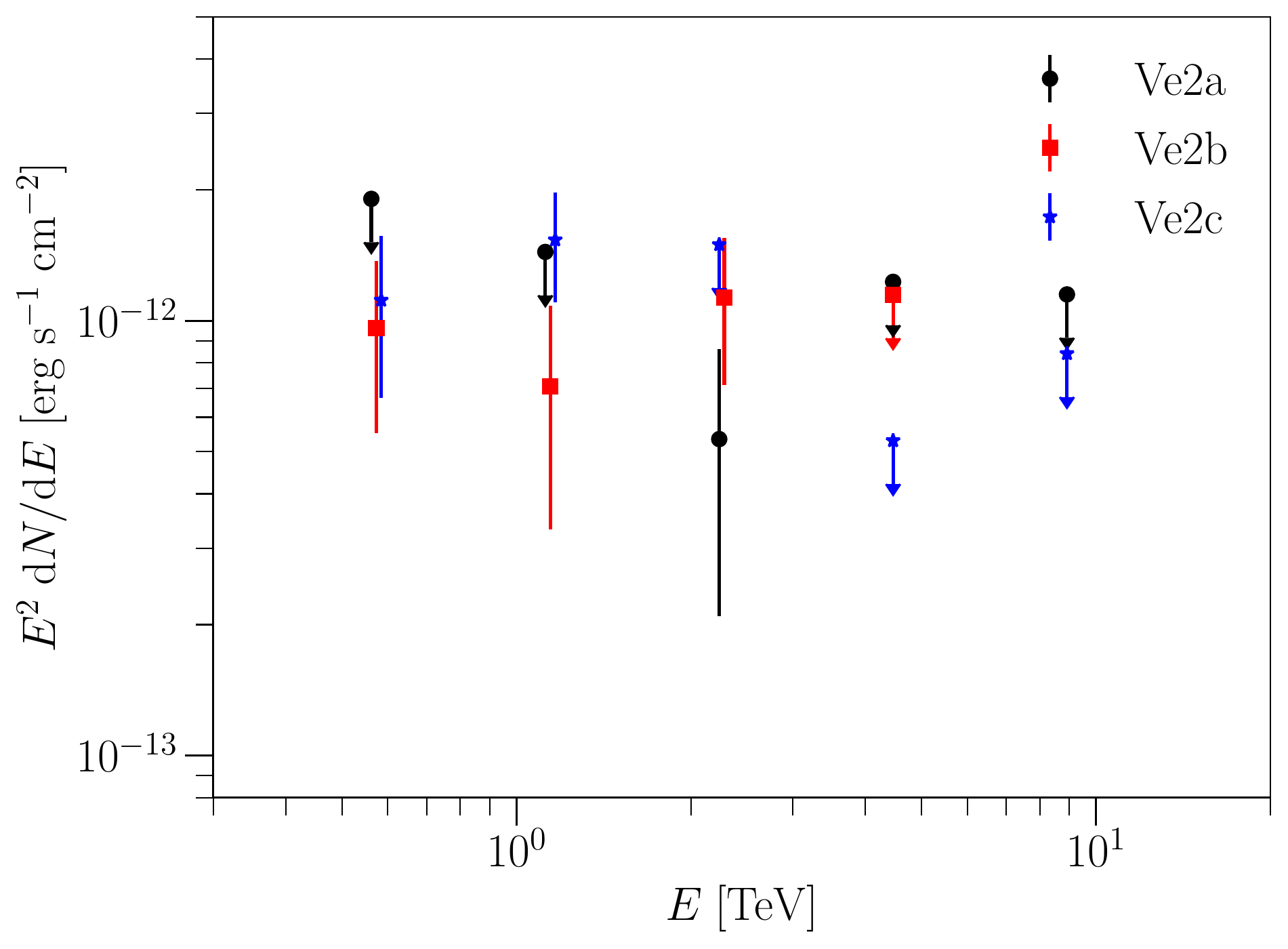}

\caption{\label{fig:data-sed-vts}
SED derived from VERITAS observations. Upper plot shows the 2017 data (Ve1a and Ve1b) and bottom plot shows 2019/20 data (Ve2a, Ve2b and Ve2c). Values of the fluxes can be found in Table~\ref{tab:veritas-results}.}

\end{figure}

Light-intensity calibration factors obtained from regular monitoring of the optical throughput and of the detector performance were applied to take time-dependent changes in the instrument response into account. The previously published result \citep{Archer-2020} did not use our newest calibration corrections which are, however, much smaller than the statistical uncertainties on that data. 
A summary of VERITAS observations and results, including the re-calibrated data from 2017 observations \citep{Archer-2020}, can be found in Table~\ref{tab:veritas-results}, and the SEDs are shown in Fig.~\ref{fig:data-sed-vts}.

\begin{deluxetable*}{c c c c c c}
 \tablecolumns{6}
    \centering
  \tablecaption{Summary of VERITAS observations and results. Fluxes are calculated above an energy threshold of 350 GeV. Upper flux limits are determined for a 95\% confidence level using the bounded method of \cite{Rolke-2005}.  Luminosities are derived from the flux values assuming a distance of 1.4 kpc.\label{tab:veritas-results}}
  
    \tablehead{
        \colhead{Obs.} & {MJD Range} & \begin{tabular}{@{}c@{}}
            \colhead{Observation Time} \\
            \colhead{(hours)}
        \end{tabular} & \begin{tabular}{@{}c@{}}
            \colhead{Significance} \\
            \colhead{($\sigma$)}
        \end{tabular} & \begin{tabular}{@{}c@{}}
            \colhead{Flux*} \\
            \colhead{($10^{-12}$ cm$^{-2}$ s$^{-1}$)}
        \end{tabular} & \begin{tabular}{@{}c@{}}
            \colhead{Luminosity} \\
            \colhead{($10^{32}$ erg s$^{-1}$)}
        \end{tabular}
         }
   \startdata
   Ve1a$^\dag$ & 58073--58083 & 7.4 & 5.7 & $2.8\pm0.6$ & $6.6\pm1.4$\\
   Ve1b$^\dag$ & 58101--58103 & 6.0 & 6.4 & $2.6\pm0.5$ & $6.1\pm1.2$\\
   \hline
   Ve2a & 58837--58851 & 6.9 & 1.3 & $<1.7^\ddag$ & $<4.0^\ddag$\\
   Ve2b & 58867--58878 & 7.8 & 4.5 & $1.6\pm0.4$ &  $3.8\pm0.9$\\
   Ve2c & 58897--58906 & 8.3 & 4.6 & $1.7\pm0.5$ & $4.0\pm1.2$\\
    \enddata
    \tablenotetext{*}{Assuming spectral index 2.6}
    \tablenotetext{\dag}{Same datasets presented in~\cite{Archer-2020} but using the VERITAS newest calibration (see text for details).}
    \tablenotetext{\ddag}{Upper limits.}
 \end{deluxetable*}    

% %%%%%%%%%%%%%%
% \begin{figure}
% \centering
% \includegraphics[width=0.9\linewidth]{DataVTS_2_std.pdf}

% \includegraphics[width=0.9\linewidth]{DataVTS_3_std.pdf}

% \includegraphics[width=0.9\linewidth]{DataVTS_4_std.pdf}
% \caption{\label{fig:data-sed-vts}
% SED derived from VERITAS observations. The solid lines show the result of the single power law fit and the gray band its 1$\sigma$ confidence interval. Values of the flux and slope can be found in Table~\ref{tab:veritas-results}}

% \end{figure}
% %%%%%%%%%%%%

%%%%%%%%%%%%%%
%%%%%%%%%%%%

% description of basic analysis steps
% cuts applied and background rejection
% list results (upper limits; non-detection significance)
% consider adding data
% systematiicerrors 

\subsection{MDM Spectral Analysis and Results} % Jules\

\begin{figure*}
\centering
\includegraphics[angle=0,trim=0 85 0 0, clip, width=1.1\linewidth]{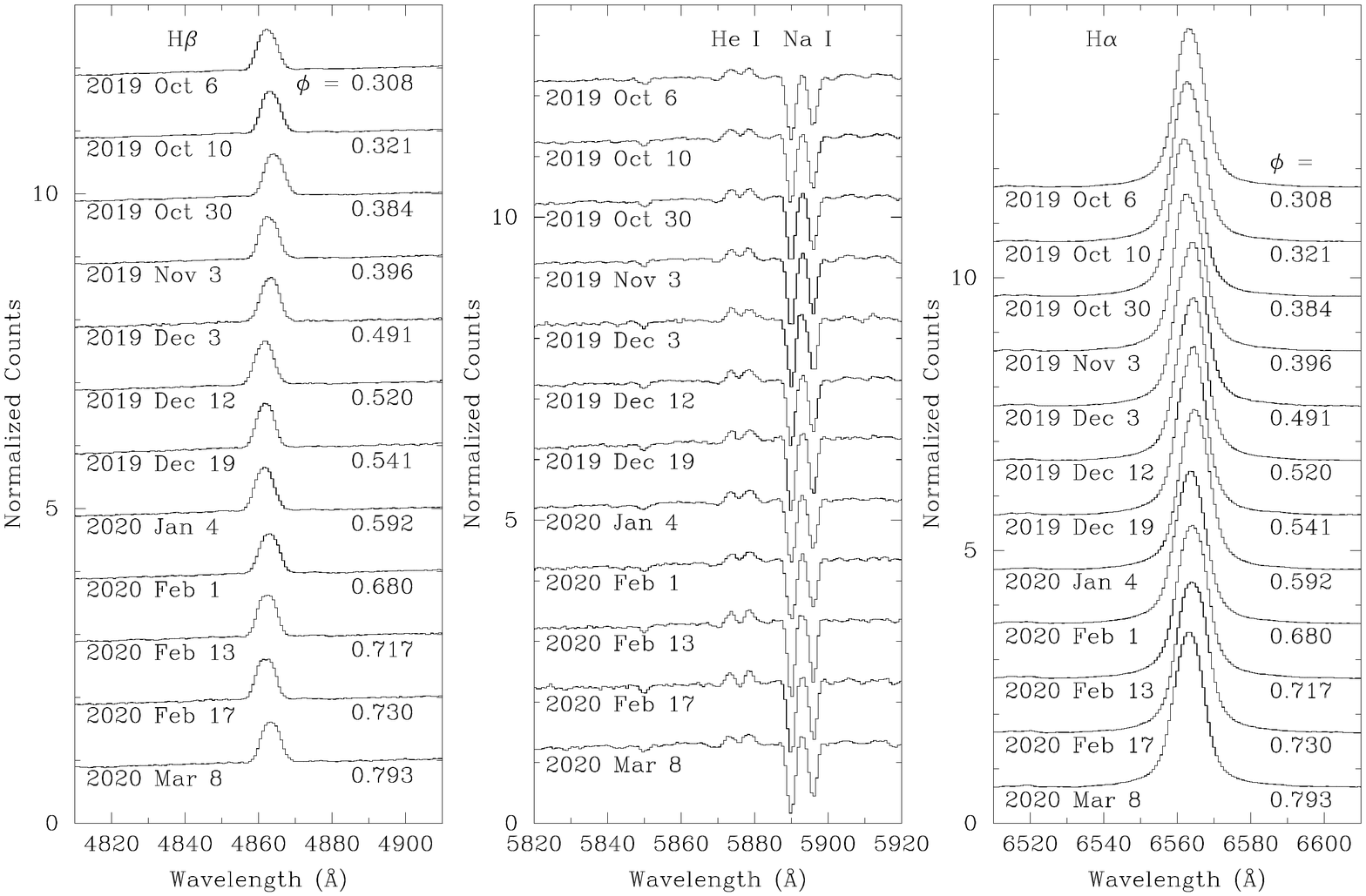}
\caption{
Selected regions of the summed optical spectra listed in Table~\ref{tab:optlog}. In each panel, the counts spectra are normalized to 1 and shifted by 1 for display purposes.  Orbital phase $\phi$ is calculated using the ephemeris of Fig.~\ref{fig:swift_lc}. Interstellar \ion{Na}{1} $\lambda\lambda5889,5895$ absorption was used as a wavelength reference. Double-peaked \ion{He}{1} $\lambda5876$ emission has a peak separation of $\approx 250$ km~s$^{-1}$.
}
\label{fig:mdm_spectra}
\end{figure*}

\label{subsec:mdm_analysis}
The optical spectra were processed and extracted using standard techniques.
Because the comparison spectra used for wavelength calibration were not taken at the same time and pointing position of the target, we added a shift in the wavelength scale to compensate for known effects of instrument flexure.
This was accomplished using the interstellar \ion{Na}{1}~D absorption-line doublet in the spectrum of the star (see Fig.~\ref{fig:mdm_spectra}) as a wavelength standard.
By shifting a spectrum to the rest frame of the interstellar feature, effects of instrument flexure, target placement within the slit, and Earth motion are removed to first order.
However, second-order effects such as temperature dependence of dispersion may be present, which would bias the wavelengths of the principal features of interest, namely the Balmer lines.
Therefore, we do not attempt to measure absolute radial velocities.
As the observing technique was not conducive to absolute spectrophotometry due to variable seeing and the narrow slit used (to preserve spectral resolution), as well as episodes of non-photometric weather, we also did not carry out flux calibration.

\begin{figure}
\centering
\includegraphics[angle=270, trim=40 30 320 500, clip, width=0.8\linewidth]{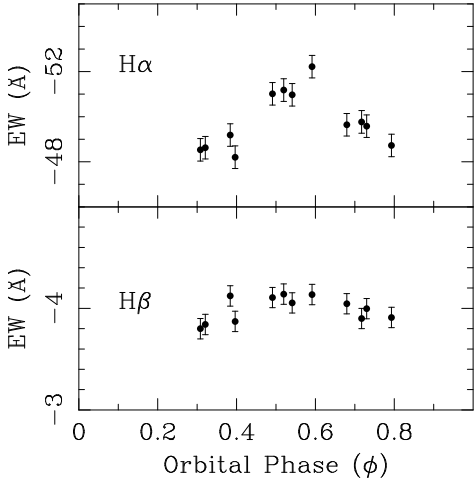}
\caption{
Equivalent widths of H$\alpha$ and H$\beta$ emission lines from Table~\ref{tab:optlog}.  Orbital phase $\phi$ is calculated using the ephemeris of Fig.~\ref{fig:swift_lc}.
}
\label{fig:mdm_ew} 
\end{figure}

The results reported here consist primarily of Balmer emission-line equivalent widths (EWs).
We first measured the EWs in individual spectra to get an estimate of their uncertainties from the variance on a given night.
Uncertainties for strong emission features such as these are dominated by systematic effects of continuum placement, and possible underlying absorption from the stellar photosphere or envelope.
Here we defined the continuum points by straight lines from 4855--4871~\AA\ for H$\beta$, and 6530--6600~\AA\ for H$\alpha$.
Typical variance among individual measurements is $\pm 0.1$~\AA\ for H$\beta$ and $\pm 0.5$~\AA\ for H$\alpha$.
Then we measured the summed spectra for each night (shown in Fig.~\ref{fig:mdm_spectra}), with results listed in Table~\ref{tab:optlog}.
There is small but significant variation in H$\alpha$ EW, which ranges from --48.2~\AA\ to --52.2~\AA, and possible correlated variability in H$\beta$.
Stronger lines  were measured in 2019 December and 2020 January than earlier or later. 
These values are also graphed as a function of orbital phase $\phi$ in Fig.~\ref{fig:mdm_ew}.

\begin{deluxetable*}{lccccc}
\tablewidth{0pt}
\tablecolumns{6}
\tablecaption{Summary of EW and orbital phase for optical spectra.}
\tablehead{
  \colhead{Date} & \colhead{Date} & \colhead{Exposure} & \colhead{H$\beta$ EW\tablenotemark{a}} &
  \colhead{H$\alpha$ EW\tablenotemark{a}} & \colhead{Phase\tablenotemark{b}} \\
  \colhead{(UT)} & \colhead{(MJD)} & \colhead{(s)} & \colhead{(\AA)} &
  \colhead{(\AA)} & \colhead{($\phi$)}
}
  \startdata
  2019 Oct 6   &  58762.43  &  $3\times60$  &  $-3.80$  &  $-48.5$  &  0.308 \\
  2019 Oct 10  &  58766.42  &  $3\times60$  &  $-3.84$  &  $-48.6$  &  0.321 \\
  2019 Oct 30  &  58786.40  &  $3\times40$  &  $-4.12$  &  $-49.2$  &  0.384 \\
  2019 Nov 3   &  58790.36  &  $3\times40$  &  $-3.87$  &  $-48.2$  &  0.396 \\
  2019 Dec 3   &  58820.40  &  $3\times40$  &  $-4.11$  &  $-51.0$  &  0.491 \\
  2019 Dec 12  &  58829.48  &  $3\times30$  &  $-4.14$  &  $-51.2$  &  0.520 \\
  2019 Dec 19  &  58836.31  &  $3\times30$  &  $-4.05$  &  $-51.0$  &  0.541 \\
  2020 Jan 4   &  58852.31  &  $3\times40$  &  $-4.14$  &  $-52.2$  &  0.592 \\
  2020 Feb 1   &  58880.21  &  $3\times40$  &  $-4.05$  &  $-49.6$  &  0.680 \\
  2020 Feb 13  &  58892.19  &  $3\times40$  &  $-3.90$  &  $-49.8$  &  0.717 \\
  2020 Feb 17  &  58896.22  &  $3\times40$  &  $-4.00$  &  $-49.6$  &  0.730 \\
  2020 Mar 8   &  58916.22  &  $3\times40$  &  $-3.91$  &  $-48.7$  &  0.793 
\enddata
\tablenotetext{a}{Typical uncertainty is $\pm0.1$~\AA\ for H$\beta$ and $\pm0.5$~\AA\ for H$\alpha$ (see text).}
\tablenotetext{b}{Orbital phase defined by the ephemeris of Fig.~\ref{fig:swift_lc}.}
\label{tab:optlog}
\end{deluxetable*}

The pattern of variability in EW closely matches that observed by \citet{Casares-2012,cas12b} from 2010 September to 2011 May.  In that period, the H$\alpha$ EW varied between $-50$~\AA\ and $-56$~\AA\, with a maximum at $\phi=0.5$.  Similarly, \citet{Aragona-2010} observed an EW of $-52.3$~\AA.  Historically, these are the largest values observed.  But at other epochs the emission line has been weaker at the same orbital phases, for example, EW$=-44$~\AA\ in 2018 \citep{sto18a,sto18b}, EW$=-30$~\AA\ in 2013--2014 \citep{Moritani-2015,Zamanov-2016}, and EW$=-14$~\AA\ in early 2010 \citep{Casares-2012,cas12b}, indicating a smaller circumstellar disk at those times.  Thus, the changes in emission-line EW cannot be entirely due to orbital dependence of the interaction of the compact object with the disk, but must be intrinsic to the Be star.  

In addition to the small variation in equivalent width over the orbit, the H$\alpha$ line profile shows changes in skewness (Fig.~\ref{fig:mdm_spectra}) like that observed by \citet{Aragona-2010} and \citet{Moritani-2015} at higher spectral resolution, which those authors attributed to waves excited in the disk by the compact object. These are not to be interpreted as the orbital radial velocity of the Be star.  In contrast, the double peaks of the \ion{He}{1}~$\lambda5876$ emission line remain stable in wavelength, as seen in Fig.~\ref{fig:mdm_spectra}, probably because this feature comes from smaller (unperturbed) radii in the disk than does H$\alpha$. (The double peaks represent the rotation velocity of the outer part of the emitting area of the disk.  Thus H$\alpha$ emission, in which peaks are not resolved, encompass larger radii, which have lower rotation velocity.)

%%%%%%%%%%%%%%%%%%%%%%%%%%%%%%%%%%%%%%%%%%%%%%%%%%%%%
\section{X-Ray light curve}
\label{sec:lc}
\subsection{X-Ray light curve model}% Hongjun
\label{subsec:lc_model}
In the pulsar scenario for TGBs \citep[][]{Dubus-2006}, it is assumed that modulated X-ray emission originates primarily in an intra-binary shock (IBS).  Interaction between the pulsar and companion winds forms a hollow cone-shaped IBS \citep[e.g.,][]{Kandel-2019}, where pulsar wind particles are accelerated to high energies \citep[e.g.,][]{Bosch-Ramon-2012}.
The accelerated particles then emit photons via synchrotron \citep[X-rays;][]{Tavani-1997} and inverse Compton scattering (ICS) processes which modulate orbitally due to the variation of the viewing angle and ICS geometry \citep[e.g.,][]{Dubus-2015}.

In this scenario, the TeV light curve is determined by ICS radiation and $\gamma$-$\gamma$ absorption, and is sensitive to the electron spectrum, orbital geometry, and seed photon density (e.g., the stellar emission), which are not yet well-known \citep[e.g.,][]{Malyshev-2019}.
We therefore focus on the X-ray light curve (Fig.~\ref{fig:swift_lc}), which is determined mostly by the orbital geometry.
The observed light curve exhibits a spike at phase $\sim$0.3 and a broad bump at phase $\sim$0.7, and is similar to the light curve of the TGB 1FGL~J1018.6$-$5856 \citep[e.g.,][]{Ackermann-2012}.
Such double-peaked light curves cannot be explained with a single component IBS, and thus we use a two-component IBS model \citep[e.g.,][]{An-2017}.

Our IBS model assumes two populations of electrons: one that moves slowly in the shock with the bulk Lorentz factor $\Gamma_D\approx 1$ and another that is bulk-accelerated along the flow with $\Gamma_D$ increasing to a maximum value ($\Gamma_{\rm D,max}$) along the flow as is predicted in hydrodynamic simulations \citep[e.g.,][]{Bogovalov-2008,Dubus-2015}.
When the observer's line of sight (LoS) crosses the tangent of the shock (e.g., the surface of the hollow cone), the observer sees an increase of the IBS emission due to beaming (a peak in the light curve).
In addition, if the orbit is eccentric, the magnetic field at the shock ($B$), which is assumed to be provided by the pulsar, modulates because of varying distance from the pulsar to IBS, as does the synchrotron emission.
Hence the two peaks in the light curve are easily explained in this model; one near the periastron (large $B$) and the other near the pulsar inferior conjunction (strong beaming).
This model successfully explained the double peak light curve of 1FGL~J1018.6$-$5856 \citep[][]{An-2017}.

%%%%%%%%%%%%%%%%%%%%%%%%%%%%%%%%%%%%%%%%%%%%%%%%%%%%%
\subsection{X-Ray light curve results} % Hongjun
\label{subsec:lc_results}
%%%%%%%%%%%%%%
\begin{figure}
\begin{tabular}{c}
\includegraphics[angle=0,width=85mm]{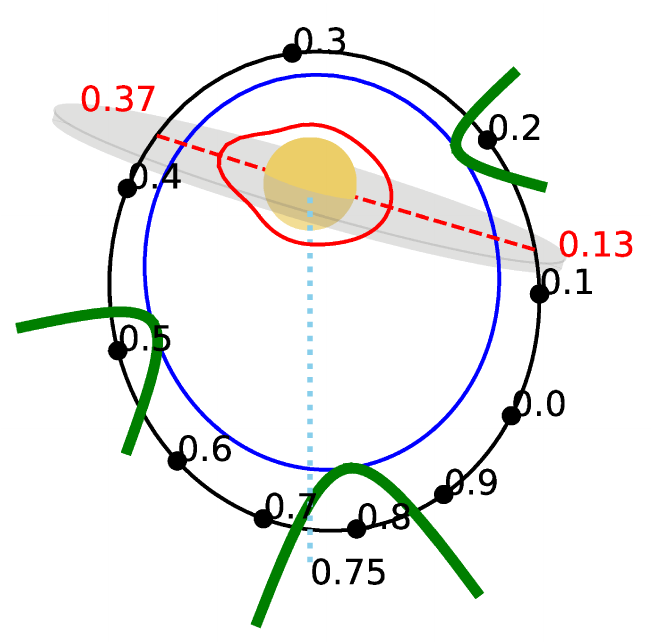} \\
\includegraphics[angle=0,width=85mm]{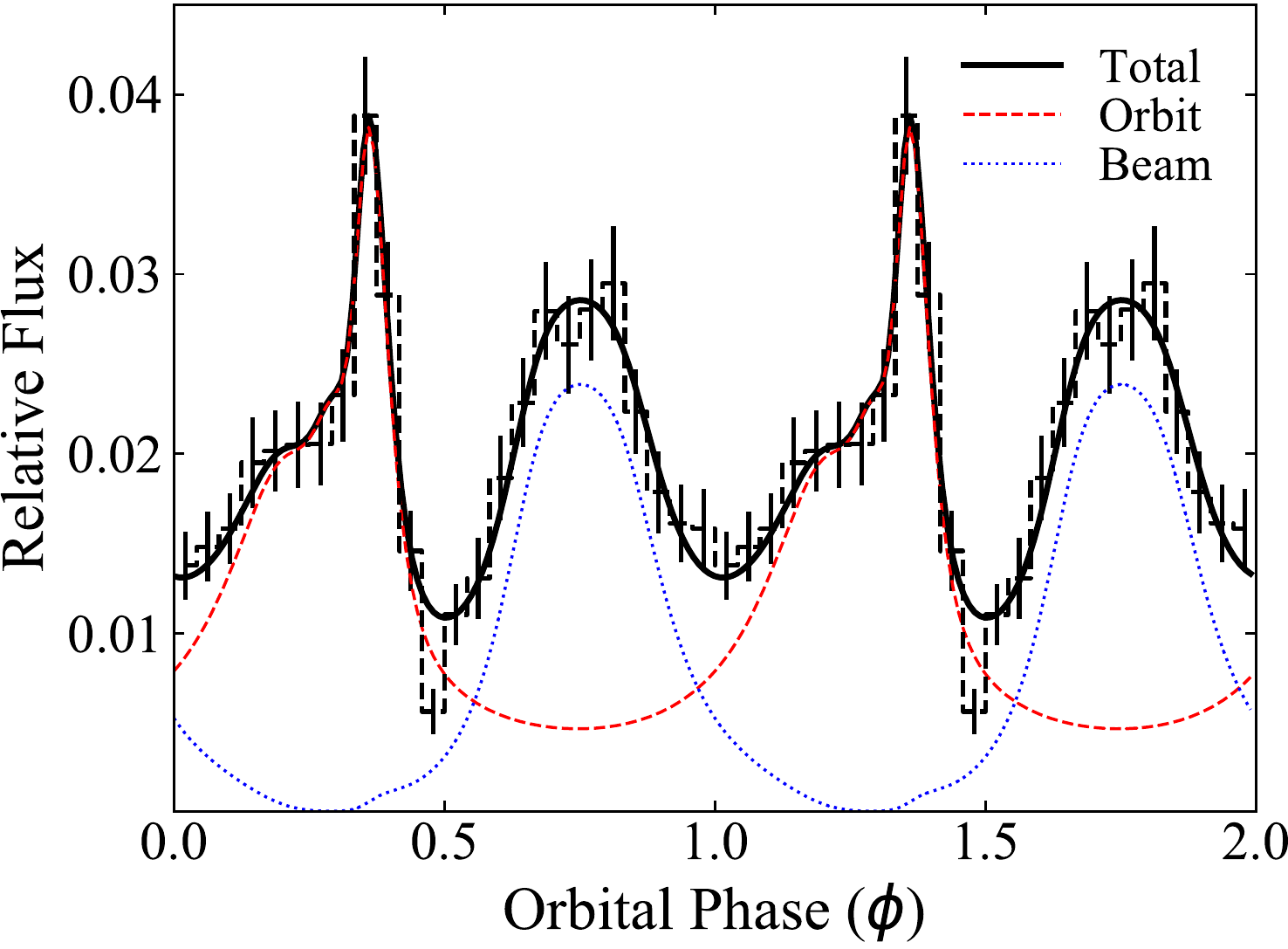} \\
\end{tabular}
\caption{\label{fig:lc_model} Top: Assumed orbit of the pulsar (black) and location of the shock nose (blue).  The relative magnetic field strength at the nose as a function of phase is indicated in red by radial distance from the pulsar. The cyan dotted line shows the observer LoS, and the cross section of the inclined disk is indicated by the red dashed line. Green curves illustrate the cross sections of the IBS at a few phases for reference. Bottom: A binned X-ray light curve (data points) and light-curve model (black). Each model component (red: orbit+disk, blue: beaming) is also displayed.}
\end{figure}
%%%%%%%%%%%%

The basic shape of the X-ray light curve (Fig.~\ref{fig:swift_lc}) already suggests that the periastron should be at phase $\phi_0 \sim 0.3$ to produce the primary peak (i.e., high orbital speed), and beaming should be responsible for the secondary peak at phase $\sim$0.7, corresponding to the pulsar inferior conjunction ($\phi_{\rm IFC}$).
While this simple scenario explains the two main features of the light curve, the excess counts at phase 0--0.3 and the dip at phase $\sim$0.4 cannot be reproduced.
We therefore add an additional model component representing the passage of the pulsar through the Be star equatorial disk.

Fig.~\ref{fig:lc_model} shows the proposed orbital geometry and X-ray light-curve model.
In the model, the companion wind is assumed to be stronger than the pulsar wind by a factor of $\eta \equiv \dot M v_{\rm wind} c/\edot\sim20$, where $\dot M$ is the mass loss of the star, $v_{\rm wind}$ is the velocity of the stellar wind, and $\edot$ is the pulsar spin-down power.
Thus the shock wraps around the pulsar. %($\dot E_{\rm sd}$ ) \dot E_{\rm sd}
The LoS is at phase $\phi_{\rm IFC}=0.75$ where the observer sees the tail of the shock, and beamed emission produces the secondary peak.
The periastron is at phase $0.3$ and the companion's disk intersects the orbit at phases $\phi_{D,1}=0.13$ and $\phi_{D,2}=0.37$.
The role of the disk for the IBS emission is rather unclear since we do not know the properties of the disk well.
In order to match the observed light curve, we can increase the magnetic field at the nose of the shock by the amount of $B_D(\phi)= B_{D0}e^{-(\phi-\phi_{D,i})^2/(2\sigma_{D,i}^2)}$, where $B_{D0}$ is the amplitude of the additional magnetic field (Table~\ref{ta:lc_model}), $\phi$ is the phase of the pulsar, and $\phi_{D,i}$ are the phases of disk-crossing. Note that this increase in $B$ is phenomenological, and the additional model component can also be achieved by modulating other covarying parameters, such as the particle density.

\begin{table}[t]
\vspace{-0.0in}
\begin{center}
\caption{Parameters for the light-curve model in Fig.~\ref{fig:lc_model}}
\label{ta:lc_model}
\vspace{-0.05in}
\scriptsize{
\begin{tabular}{lccc} \hline\hline
Parameter       & Symbol        & Value     \\ \hline
Semi-major axis (cm) & $a$ & $3.9\times 10^{13}$ \\
Eccentricity & $e$ & 0.45 \\
Inclination (deg.)   & $i$ & 47 \\
Periastron phase & $\phi_0$ & 0.3 \\ 
Pulsar inferior conjunction phase & $\phi_{\rm IFC}$ & 0.75 \\ \hline
Wind momentum-flux ratio  & $\eta$ & 22 \\
Magnetic field (G)$^a$   & $B$   & 1.5 \\
Max. bulk Lorentz factor$^b$  & $\Gamma_{\rm D,max}$      & 7 \\
Electron spectral index & $p_1$ & 2.3 \\
Min. electron Lorentz factor    & $\gamma_{\rm e, min}$ & $7\times 10^3$ \\
Max. electron Lorentz factor    & $\gamma_{\rm e, max}$ & $3\times 10^7$ \\ \hline
Disk crossing phases & $\phi_{D,1},\phi_{D,2}$ & 0.13, 0.37  \\
Projected Disk width (phase angle, deg.) & $\sigma_{D,1}$, $\sigma_{D,2}$ & 29, 18  \\ 
Magnetic field at disk interaction (G)   & $B_{D0}$   & 0.6, 0.9 \\
\hline
\end{tabular}}
\end{center}
\vspace{-0.5 mm}
\footnotesize{
$^{\rm a}$ At the shock nose at $\phi_{\rm IFC}$.\\
$^{\rm b}$ Increases linearly with travel distance for the fast population.\\}
\end{table}

In our construction of the orbit and disk, the sharp increase in the column density $N_{\rm H}$ at the primary peak $\phi_{D,2}=0.37$ \citep[e.g.,][]{Moritani-2015,Malyshev-2019} is easily explained as being due to enhanced absorption by dense material in the disk, i.e., X-ray emission originates from within the disk and propagates through it nearly ``edge-on'' along the LoS \citep[e.g.,][]{Aragona-2010}. 

The small excess at phase $\sim 0.15$ can be explained by the pulsar-disk interaction on the other side of the orbit ($\phi_{D,1}=0.13$; Fig.~\ref{fig:lc_model}); evidence for an $N_{\rm H}$ increase is observed around this phase (see Sec.~\ref{subsec:jointfit}).
%perhaps because of a rapid density drop $n\propto r^{-3}$ in the disk \citep[e.g.,][]{Klement-2017} and shorter path length along the LoS.
The reconstructed X-ray light curve is shown in Fig.~\ref{fig:lc_model} and model parameters are presented in Table~\ref{ta:lc_model}.
Note the strong parameter covariance, so that other sets of parameters, especially those related to the IBS and electrons (e.g., $B$, $\gamma_e$'s, $p_1$ etc), can equally well explain the light curve.
However, the phases for the periastron, disk crossings, and the inferior conjunction are robustly determined in this modeling.

%%%%%%%%%%%%%%%%%%%%%%%%%%%%%%%%%%%%%%%%%%%%%%%%%%%%%

\subsection{Modeling Disk Passage With \nustar/\swift Joint Spectral Fitting} % Yarone
\label{subsec:jointfit}

We expect pulsar passage through the disk to show up in our spectra as an increase in \nh.
The first disk passage in our proposed orbital solution occurs at $\phi_{D,1} = 0.13$, which fortuitously coincides with the date of Nu1a (2017-11-22; $\phi\sim0.14$), and its simultaneous \swift observation (``Sw1a'': obsID 00088078001; 6.75 ks; $\sim145$ source counts).
A significant drawback of attempts to find indications of disk disruption in soft X-ray spectra alone \citep[e.g.,][]{Malyshev-2019} is that there is degeneracy between $\Gamma$ and \nh.
However, since the effects of absorption are negligible above 3.0 keV, Sw1a (0.2--5.6 keV) and Nu1a (3.0--30.0 keV) together present a unique opportunity to further probe our light curve model by looking for an increase in \nh compared with observations of \srca at other phases.

\swift spectra and response files were generated with \texttt{XSELECT} (v2.4k) and the \texttt{xrtmkarf} task, using the source and background regions described in Sec.~\ref{subsec:swift_analysis}.
All spectra (Sw1a, Nu1a fpmA, and Nu1a fpmB) were grouped to a minimum of 20 counts per bin and jointly fit to an absorbed powerlaw, with the photon index frozen to $\Gamma = 1.77$ (Sec.~\ref{subsec:nustar_analysis}).
The fit yielded a column density of $(0.59^{+0.35}_{-0.32}) \times10^{22}$ cm$^{-2}$ (90\% conf.; red.~$\chi^2 = 0.96$, 95 dof).
Meanwhile, the same analysis applied to Nu1b (2017-12-14, $\phi\sim 0.22$; $\Gamma = 1.56$), with its simultaneous \swift observation (``Sw1b'': obsID 00088078002; 7.10 ks; $\sim117$ source counts) yielded \nh = $(0.32^{+0.50}_{-0.32}) \times10^{22}$ cm$^{-2}$ (red.~$\chi^2 = 0.83$, 94 dof).
Simultaneous \swift observations for Nu2a and Nu2b (``Sw2a'' and ``Sw2b'') had insufficient counts to perform a similar analysis (28 and 41 source counts, respectively).
The median value of previously reported \srca \nh measurements \citep[e.g.,][]{Malyshev-2019} is 0.30$\times10^{22}$ cm$^{-2}$.

The joint spectral fit is suggestive, but the error bars on \nh are too large to indicate an increase in column density at $\phi_{D,1}$.
This is expected due to the paucity of \swift counts relative to \nustar.
In principle, the spectra from multiple co-phased \swift observations can be combined to mitigate this problem, but as indicated in Sec.~\ref{subsec:mdm_analysis}, the Be star disk likely has inherent variability (in addition to disruptions due to disk passage by the pulsar), resulting in super-orbital modulation, so observations would need to be from the same orbit in order to expect agreement between spectra.
% since they share an instrument response (\texttt{.rmf}) file, 
% A shortcoming of the above analysis is that, due to the paucity of \swift counts relative to \nustar, the error bars on \nh are too large to definitively distinguish Sw1a from other reported values.}

\begin{figure}
\includegraphics[angle=0,width=85mm]{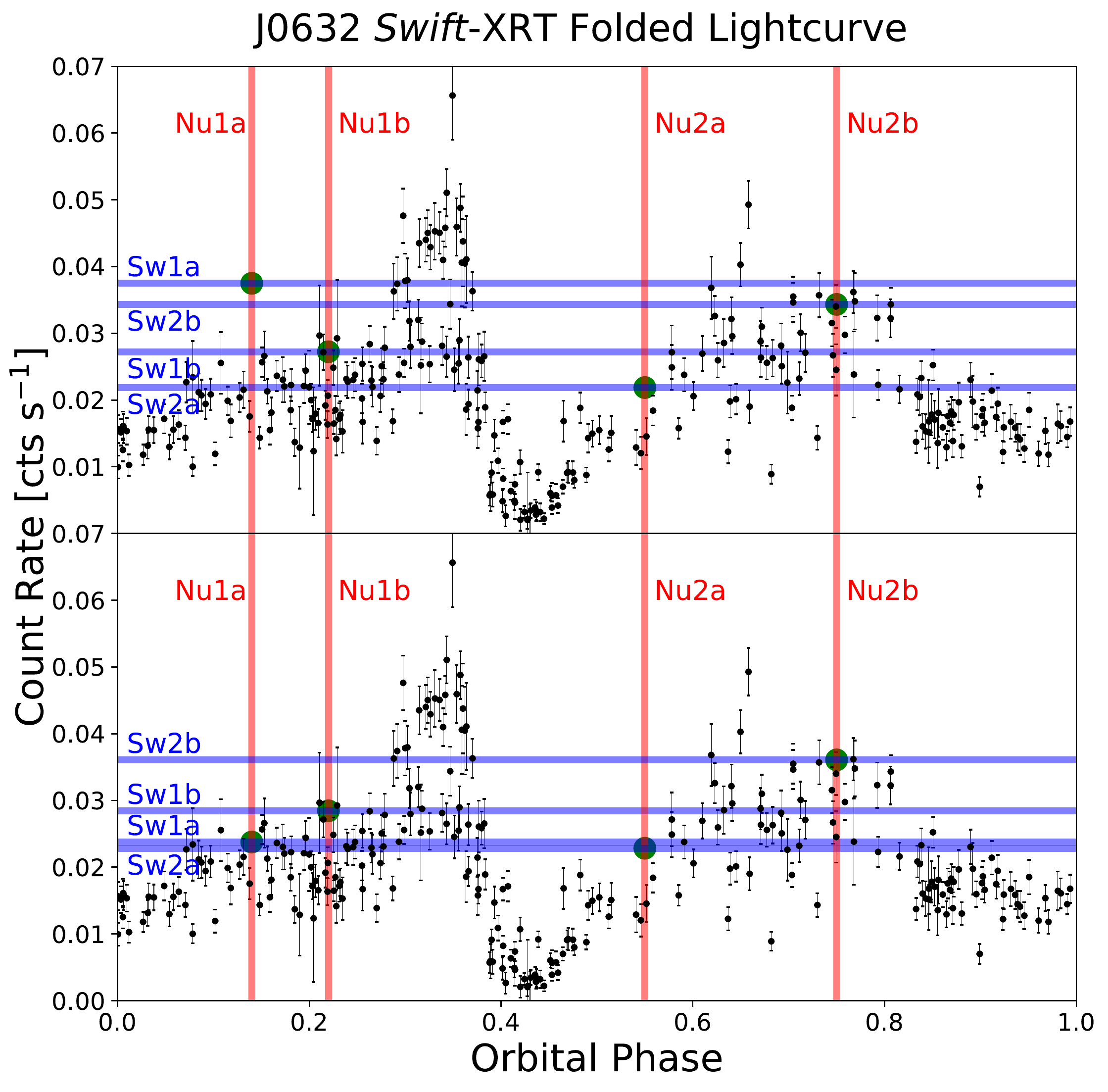}
\caption{\label{fig:pimms_results}
\texttt{PIMMS} prediction of \swift count rates for each \nustar observation, overlaid onto the folded light curve.  Vertical red lines correspond to the phases of the \nustar observations, and horizontal blue lines indicate the predicted \swift count rates.  Intersections for corresponding observations are marked with a green dot.\\ 
Top: \nh $= 0.30\times10^{22}$ cm$^{-2}$ for all \texttt{PIMMS} estimates\\
Bottom: Adjusted \texttt{PIMMS} estimate for Sw1a using \nh = $1\times10^{22}$ cm$^{-2}$.}

\end{figure}

In order to further explore the possibility of an \nh spike, we used NASA's Portable Interactive Multi-Mission Simulator (\texttt{PIMMS}) tool to simulate \swift 0.2--10.0 keV count rates based on the 3.0--20.0 keV fluxes of our \nustar observations.
% Due to the paucity of counts in Sw1a, we could not precisely determine \nh at that phase by simply using Nu1a as a lever arm to constrain $\Gamma$ and then fitting Sw1a.
% Instead, in order to look for consistency between our observed \nustar fluxes and the \swift light curve at the proposed orbital period, we used NASA's Portable Interactive Multi-Mission Simulator (\texttt{PIMMS}) tool to simulate \swift 0.2--10.0 keV count rates based on the 3.0--20.0 keV fluxes of our \nustar observations (Table~\ref{tab:nustar-spectral}).
We passed as additional inputs the best-fit photon indices and \nh = 0.30$\times10^{22}$ cm$^{-2}$.
Interestingly, the predicted \swift count rates based on Nu1b, Nu2a, and Nu2b were all consistent (within 1$\sigma$ errors) with the \swift light curve, while the predicted count rate from Nu1a was higher by a factor of $\sim2$.
This is shown in the top panel of Fig.~\ref{fig:pimms_results}.

We then modeled the disk passage at Nu1a as an increase in the column density to \nh = $1\times10^{22}$ cm$^{-2}$ (within 2$\sigma$ of the estimated value from the joint \swift-\nustar fit), keeping the \nustar flux and photon index the same.
%Since our orbital solution includes a disk passage at the orbital phase of Nu1a, we model that as an increase in the column density to \nh = $1\times10^{22}$ cm$^{-2}$, keeping the \nustar flux and photon index the same.
This amounts to a $\sim3$-fold increase in \nh during passage through the disk.
This estimated value of \nh is consistent with Eq.\ 2 of \cite{Klement-2017}, assuming the disk is primarily composed of hydrogen and an average photon path length of $\sim3.9\times10^{13}$ cm, the orbital separation at $\phi_{D,1}$.
%\REDCOM{Following Eq. 2 of \cite{Klement-2017}, assuming the disk is primarily composed of hydrogen, and an average photon path length of $\sim4\times10^{11}$ cm, we estimate a hydrogen column density through the disk to be $\sim4.3\times10^{21}$ cm$^{-2}$.
%Adding this to our median value, we set \nh $\approx 1\times10^{22}$ cm$^{-2}$ for Sw1a, which amounts to a $\sim3$-fold increase in \nh during passage through the disk}.
%setting \nh = $1\times10^{22}$ cm$^{-2}$, keeping the \nustar flux and photon index the same.
The new \texttt{PIMMS} results are consistent with the folded light curve, as shown in the bottom panel of Fig.~\ref{fig:pimms_results}.
These results are suggestive of a spike in column density at $\phi_{D,1}$.

%%%%%%%%%%%%%%
Structural changes to the outer disk of MWC 148 have been indicated by changes to the H$\alpha$ emission lines \citep{sto18b}.
Thus, high resolution optical spectroscopy aimed at $\phi_{D,1}$ and $\phi_{D,2}$ may also be utilized to find signatures of disk passage.  As previously mentioned, super-orbital modulation of the circumstellar disk presents a challenge to detecting disk-passage, since we cannot be sure whether an increase in \nh (or, for that matter, changes to the H$\alpha$ emission lines) are due to disk passage or other changes to the disk structure.
If, over multiple orbits, a change is detected at the same phase as an H$\alpha$ emission line change, then that would strongly favor the disk-passage scenario.

%%%%%%%%%%%%%%%%%%%%%%%%%%%%%%%%%%%%%%%%%%%%%%%%%%%%%
\section{Interpretation of the \nustar and VERITAS SED data} % Raul
\label{sec:sed}

\begin{figure}
\centering
\includegraphics[width=0.95\linewidth]{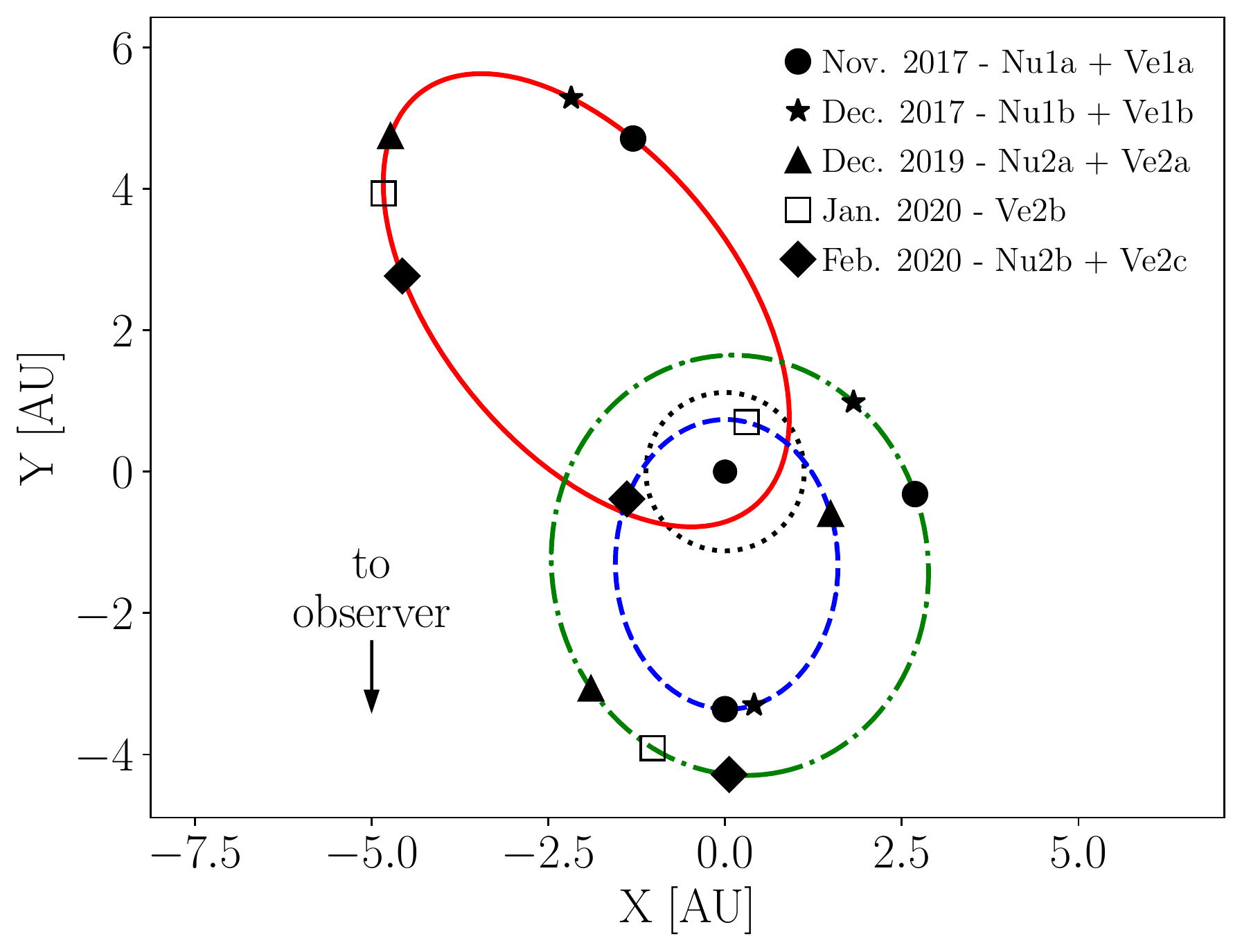}
\caption{\label{fig:sed-orbits} Illustration of the orbit of the compact object projected onto the orbital plane for the solutions from~\cite{Casares-2012} (solid red line) and~\cite{Moritani-2018} (dashed blue line) and from Sec.~\ref{sec:lc} (dash-dotted green line). See~\cite{Archer-2020} for the complete list of system parameters for the former two. The locations of the compact object during the combined \nustar and VERITAS observations of the 2017 and 2019/20 campaign are indicated as black markers. The companion star is assumed to be in a fixed position and the estimated size of the circumstellar disk~\citep{Moritani-2015, Zamanov-2016} is indicated by a dotted black line.}
\end{figure}

In this section we provide a possible interpretation for the SED data derived from the contemporaneous \nustar and VERITAS observations. The procedure will follow the previous work of~\cite{Archer-2020}, where the SED data from the 2017 campaign were well described by a one-zone leptonic model, based on the pulsar scenario. The basic assumptions of the model are already described in Sec.~\ref{sec:lc}. However, unlike in the previous section, the assumption here is that only one population of high-energy electrons, distributed around the apex of the pulsar-wind shock, is responsible for the non-thermal emission. Besides the two orbital solutions by~\cite{Casares-2012} and~\cite{Moritani-2018} previously considered in~\cite{Archer-2020}, in this section we also test the orbital solution obtained in Sec.~\ref{sec:lc}. The three orbital solutions are illustrated in Fig.~\ref{fig:sed-orbits}.

%%%%%%%%%%%%%%%%%%%%%%%%%%%%%%%%%%%%%%%%%%%%%
%%%%%%%%%%%%%%%%%%%%%%%%%%%%%%%%%%%%%%%%%%%%%

The pulsar spin-down luminosity (\edot) is a central free parameter of the model, since it directly affects the position of the IBS and the B-field intensity at the emission zone. The pulsar-wind termination shock forms approximately at the same position as the discontinuity created by the collision between pulsar and stellar wind. This position is given by the condition that the pressure from both winds balance each other~\citep{Harding-1990}. Thus, a higher \edot implies that the IBS is formed farther away from the pulsar and closer to the companion star, which influences the radiative environment by changing the $B$-field intensity and the density of the photon field that is up-scattered by electrons producing gamma-rays. Another free parameter of the model is the pulsar wind magnetization at a fixed distance from the pulsar ($\sigma_0$). We assume that the pulsar wind magnetization ($\sigma$) evolves according to the relationship $\sigma = \sigma_0\;(\rsh\;/\;3\;\textrm{AU})^{-1}$, where (\rsh) is the distance from the pulsar. For a given \edot, $\sigma$ and IBS position, the B-field intensity is calculated following~\cite{Kennel-1984a, Kennel-1984b}. 

In the context of this model, the stellar disk represents a region in which the mass loss rate associated with the stellar wind is substantially higher~\citep{Waters-1988} than the isotropic component. As a first order consequence, during the passage of the pulsar through the disk, the termination shock would form closer to the pulsar and farther from the companion star, changing the radiative environment and the non-thermal emission patterns. Because of the large uncertainties on the description of the properties and geometry of the disk (size and relative inclination)~\citep{Moritani-2015, Zamanov-2016}, our model only takes the isotropic stellar wind into account. The implications of this simplification are discussed later in this section.

The energy spectrum of the high-energy electron population for each set of SEDs is described by a power law function at the relevant energy range (0.1--5 TeV), where its spectral index is derived from the observed X-ray SED ($\Gamma_\textrm{elec} = 2\;\Gamma_\textrm{X-rays} - 1$) and its normalization is treated as a free parameter of the fit. Therefore, the modelling of the electron energy losses is not necessary, which is advantageous considering the difficulties related to model descriptions of the adiabatic processes. The full description of the model can be found in~\cite{Archer-2020}.     

Data from Nu1a, Nu1b, Ve1a, and Ve1b \citep{Archer-2020} were combined with Nu2a, Nu2b, Ve2a, and Ve2c (this paper) to perform the SED fit. Ve2b was not included because data at both bands are required. The final SED data include 4 sets from different time periods (Nov. 2017, Dec. 2017, Dec. 2019 and Feb. 2020). In Fig.~\ref{fig:sed-orbits} we show illustrations of the orbital solutions and the respective positions of the pulsar at each observation for each solution. 

The model contains 6 free parameters (the normalization of the electron spectrum for each period, \edot and $\sigma_0$). To calculate the SED curves, we used the Naima package~\citep{Zabalza-2015}, which follows the calculations from~\cite{Blumenthal-1970}. The model fitting was performed by means of a $\chi^2$ method, in which \edot and $\sigma_0$ are scanned over a pre-defined grid and the electron spectra normalizations are fitted by minimizing $\chi^2$ using Minuit framework~\citep{James-1975}.  

\begin{figure*}
\centering
\includegraphics[width=0.95\textwidth]{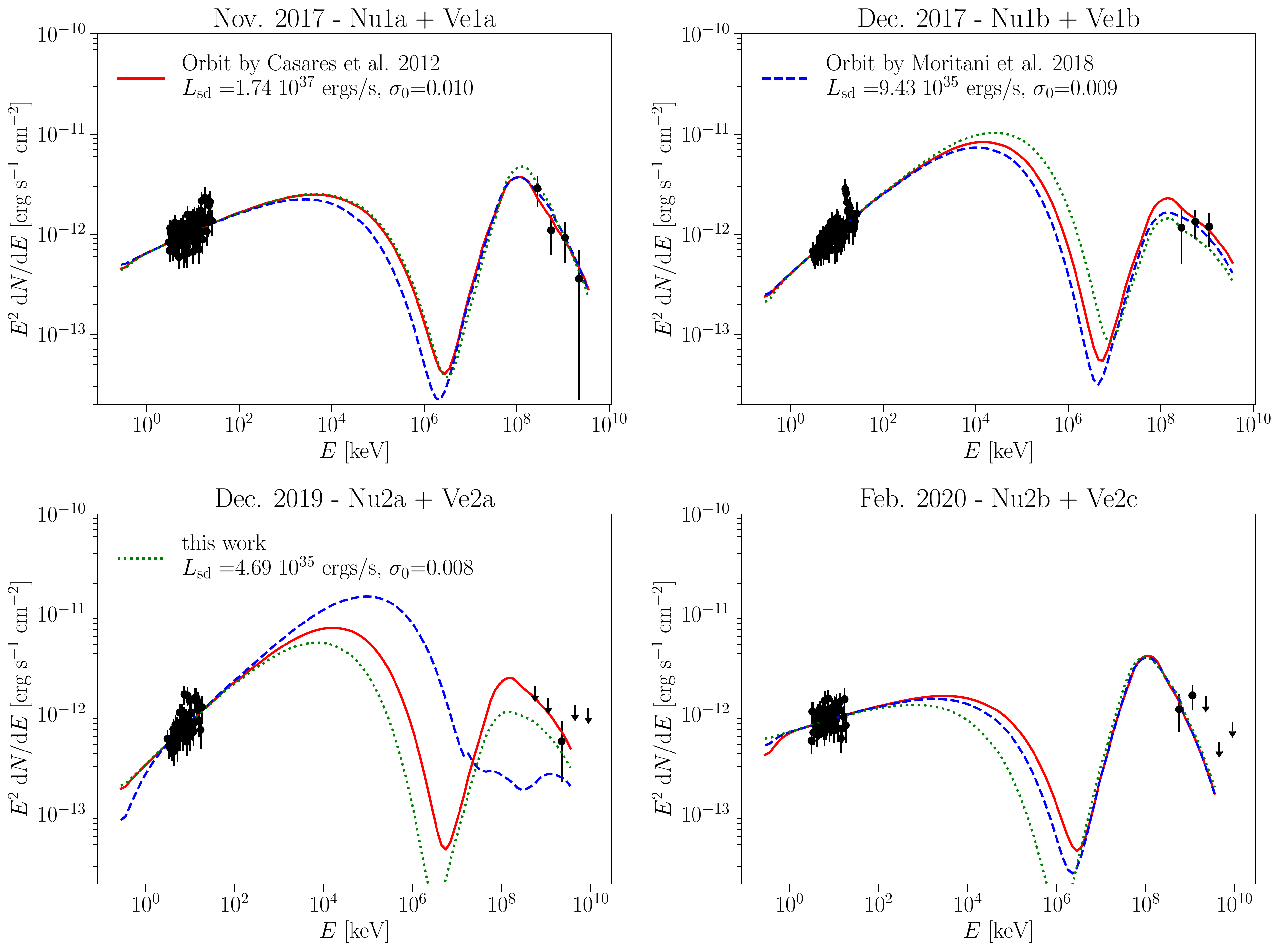}
\caption{\label{fig:sed-model} SED data-model comparison assuming the best solution of the model fitting for the orbital solutions from ~\cite{Casares-2012} (solid red line), \cite{Moritani-2018} (blue dashed line) and from Sec.~\ref{sec:lc} (green dotted line).}
\end{figure*}

The best fit solutions were found at ($\edot=1.74\times10^{37}$ ergs/s, $\sigma_0=0.010$) for the~\cite{Casares-2012} orbital solution, at ($\edot=9.43\times10^{35}$ ergs/s, $\sigma_0=0.009$) for the~\cite{Moritani-2018} one and ($\edot=4.69\times10^{35}$ ergs/s, $\sigma_0=0.008$) for the solution obtained in Sec.~\ref{sec:lc}. The SEDs' model-data comparison for these fit solutions are shown in Fig.~\ref{fig:sed-model}. In order to illustrate the whole energy range, the electron energy spectra were assumed to start at 0.1 GeV, and a exponential cutoff was added at 5 TeV. Note that these values are arbitrary and the performed fit is not sensitive to them.

%%%%%%%%%%%%%%%%%%%%%%%%%%%%%%%%%%%%%%%%%%%%%
%%%%%%%%%%%%%%%%%%%%%%%%%%%%%%%%%%%%%%%%%%%%%

The results obtained here are consistent with those obtained in~\cite{Archer-2020} with a smaller data set covering the rising of the first light-curve peak, showing that our one-zone leptonic model describes well the contemporaneous \nustar and VERITAS SED data. In~\cite{Archer-2020}, however, the effect of the circumstellar disk could be neglected because the relative distances between both stars were substantially larger than the extension of the disk. Since the data from the 2019/20 campaign were partially taken when the pulsar was in close proximity to the companion star (see Fig~\ref{fig:sed-orbits}), this assumption is not valid for the present analysis.
However, the fact that our simplified model, which neglects the disk, still fits the data well can have interesting implications. One possibility is that the effect of the disk at the position of the termination shock is too small to influence our model description, given that the statistical uncertainties in the VERITAS SED are relatively large. In this scenario, the correlation between the two light curve peaks with disk passages could still be explained by adiabatic energy losses, instead of variations in the $B$-field and/or ICS photon field density. 

% However, unlike in~\cite{Archer-2020}, the effect of the circumstellar disk cannot be neglected only based on the relative position of both stars. As seen in Fig~\ref{fig:sed-orbits}, the data from the 2019/20 campaign were partially taken when the pulsar was in close proximity to the companion star.

%%%%%%%%%%%%%%%%%%%%%%%%%%%%%%%%%%%%%%%%%%%%%%%%%%%%%
\section{Summary and Discussion} % Yarone
\label{sec:conclusions}

% Despite extensive observation at high energies, \srca remains an enigmatic source.
% An orbital solution that explains its distinctive light curve, and an emission model that explains its orbitally modulated spectrum, remain elusive.
% In this study, we presented a summary of all X-ray and gamma-ray observations of \srca by \nustar and VERITAS, which now span several orbital phases, as well as the first MDM observations in the optical band.  We searched for H$\alpha$ variation in the optical band, probed a new light curve model using X-ray data from the entire orbit, and then tested our orbital solution and broadband SED using hard X-ray and gamma-ray data. Each part corroborated our X-ray light curve model and its corresponding orbital solution.
We presented a multi-wavelength observation campaign of the rare TeV gamma-ray binary \src with \nustar (X-ray), VERITAS (TeV gamma-ray), and MDM (optical).  The observations took place in the secondary peak of \srca's double-peaked light curve as a follow up to our previous study, whose observations were targeted at the primary peak \citep{Archer-2020}. Signatures of disk-pulsar interactions, indicated by the light curve model of \citep{Malyshev-2019}, were not detected by MDM.

Two models were applied to the system in this study: (1) A detailed X-ray light curve model based on \cite{An-2017}, using our analysis of a decade of archival \swift data. This set forth a new orbital solution and robustly determined key orbital phases.  (2) A more simplified SED fit, which used broadband data from both \nustar-VERITAS campaigns.  This model constrained the intrinsic pulsar parameters for both our new orbital solution, as well as for two previously published solutions.

\begin{itemize}
    \item \textbf{X-ray and TeV spectral analyses} were performed using data from \nustar and VERITAS. The 3.0--20.0 keV \nustar\ spectra are well fit to an unabsorbed power-law model, as \nustar’s broadband capabilities allow it to measure the photon index with relatively high precision, and little to no degeneracy with \nh.
    % We observed a spectral variation between the two observations, from $\Gamma = 1.57\pm0.10$ to $\Gamma = 1.79\pm0.08$.
    Between the two campaigns, an inverse relationship was observed between spectral hardness and X-ray flux (see Table~\ref{tab:nustar-spectral}).
    Such spectral variability was seen in a previous study \citep[][]{Malyshev-2019} and was attributed to particle cooling, which can qualitatively explain the observed relationship.  In addition, there may be intrinsic variability in the particle injection spectrum, perhaps due to varying shock obliquity with orbital phase. No pulsation signals below 31.25 Hz were detected in the \nustar timing data.  VERITAS performed nearly simultaneous TeV observations of the source along with the two \nustar observations (between 2019, Dec 20 and 2020, Jan 3 and between 2020, Feb 18 and 28) and another observation between 2020, Jan 19 and 30. The second and the third VERITAS observations yielded $>4\sigma$ detection of the source above an energy threshold of 350 GeV, while only reporting a $1.3\sigma$ significance during its first observation in the rise toward the second peak.

    % \item \textbf{\nustar} performed two observations in the secondary peak in 2019--2020. The 3.0--20.0 keV \nustar\ spectra are well fit to an unabsorbed power-law model. We observed a spectral variation between the two observations, from $\Gamma = 1.57\pm0.10$ to $\Gamma = 1.79\pm0.08$. \nustar’s broadband capabilities allow it to measure the photon index with relatively high precision, and little to no degeneracy with \nh. Between the two campaigns, an inverse relationship was observed between spectral hardness and X-ray flux (see Table~\ref{tab:nustar-spectral}).
    % \REDCOM{Such spectral variability was seen in a previous study \citep[][]{Malyshev-2019} and was attributed to particle cooling, which can qualitatively explain the observed relationship.  In addition, there may be intrinsic variability in the particle injection spectrum, perhaps due to varying shock obliquity with orbital phase.} No pulsation signals below 31.25 Hz were detected in the \nustar timing data.
    
    % \item \textbf{VERITAS} performed nearly simultaneous TeV observations of the source along with the two \nustar observations (between 2019, Dec 20 and 2020, Jan 3 and between 2020, Feb 18 and 28) and another observation between 2020, Jan 19 and 30. The second and the third VERITAS observations yielded $>4\sigma$ detection of the source above an energy threshold of 350 GeV, while only reporting a $1.3\sigma$ significance during its first observation in the rise toward the second peak, potentially due to unexpectedly reduced exposure time.
    
    \item \textbf{No significant variation in the H$\alpha$ and H$\beta$ EWs} were detected by MDM.  Observations spanning orbital phases $\sim0.3-0.8$ were originally intended to trigger the 2020 \nustar observations, since disk-pulsar interactions usually result in H$\alpha$ or H$\beta$ line profile variation.  However, as can be seen in Fig.~\ref{fig:swift_lc}, the lack of full orbital phase coverage by MDM may have missed the crossing phase $\phi_{D,2} = 0.37$ as predicted by the best model fit to the folded X-ray lightcurve data. Further optical monitoring of the source is suggested over the orbital phases ($\phi = 0.13 $ and $0.37$) where our model predicts the disk-pulsar interactions.
    
    % \item \textbf{MDM} observations of H$\alpha$ and H$\beta$ emission lines spanned orbital phases $\sim0.3-0.8$, originally intended as a triggering condition for the 2020 \nustar observations, since disk-pulsar interactions usually result in H$\alpha$ or H$\beta$ line profile variation. No significant variations in the H$\alpha$ and H$\beta$ EWs were detected. However, as can be seen in Fig.~\ref{fig:swift_lc}, the lack of full orbital phase coverage by MDM may have missed the crossing phase $\phi_{D,2} = 0.37$ as predicted by the best model fit to the folded X-ray lightcurve data. Further optical monitoring of the source is suggested over the orbital phases ($\phi = 0.13 $ and $0.37$) where our model predicts the disk-pulsar interactions.
    %, so the possibility of disk-crossing signatures in the optical band is not ruled out.
    
    % \item \textbf{Archival \swift} data spanning over 10 years were analyzed to construct the folded X-ray light curves  and determine the orbital period more accurately.  
    % %since it covers the entire orbit with significant flux modulation. 
    % Our timing analysis yielded an orbital period of 317.3 days, consistent with the solution derived by an independent \swift analysis \cite{Maier-2019} ($317.3 \pm 0.7$ days), which is the most accurate orbital period to date.
    
    \item \textbf{A new intra-binary shock model}, adapted from \cite{An-2017},  was applied to the folded \swift light curve data. Three model components were required to account for the peculiar light curve shape characterized by a narrow primary peak, sharp dip, and broad secondary peak: (1) modulation due to the orbital eccentricity, with increased X-ray flux at the periastron; (2) beamed X-ray emission at the inferior conjunction due to a bowed shock front; (3) pulsar-disk interaction at two phases. The best-fit model robustly determined phases for the periastron, inferior conjunction, and disk-crossings at $\phi_0 = 0.30$, $\phi_\text{IFC} = 0.75 $ and $\phi_{D,i} = 0.13; 0.37$, respectively.
    To probe our orbital solution, we performed a joint spectral analysis on contemporaneous \swift and \nustar observations around $\phi_{D,1}$ to find evidence of an increase in \nh.
    Since the limited bandwidth and photon counts in the \swift data did not yield precise \nh measurements, we estimated \nh values by comparing the estimated \swift count rates based on the \nustar spectral fit results (which determined $\Gamma$ well) and the actual XRT count rates.
    
    \item \textbf{An updated multi-wavelength SED fit} with \nustar and VERITAS data,  further constrained the pulsar parameters from \cite{Archer-2020}. \edot and $\sigma$ were determined for our new orbital solution, as well as for those previously published by \cite{Casares-2012} and \cite{Moritani-2018}. The derived spin-down power values varied significantly between the orbital solutions --  e.g., the \cite{Casares-2012} solution yielded a much higher spin-down power $\edot   = 1.7\times10^{37}$ erg\,s$^{-1}$. Such a high \edot value is not consistent with the lack of a pulsar wind nebula (PWN); most of the energetic pulsars with $\edot \simgt 4\times10^{36}$ erg\,s$^{-1}$ exhibit  distinctive PWNe in the X-ray band \citep{Gotthelf-2004}. 
    %The fact that no signatures of a PWN have been found near \srca casts doubt onto the orbital solution of \cite{Casares-2012}, which yields a best-fit \edot of $1.7\times10^{37}$ erg/s.  
    Meanwhile, the lower \edot values for our new orbital solution and that of \cite{Moritani-2018} are consistent with the non-detection of PWN associated with \srca. 
\end{itemize}

\section*{Acknowledgements} % Yarone
This work used data from the \nustar\ mission, a project led by the California Institute of Technology, managed by the Jet Propulsion Laboratory, and funded by NASA. We made use of the \nustar\  Data Analysis Software (NuSTARDAS) jointly developed by the ASI Science Data Center (ASDC, Italy) and the California Institute of Technology (USA).
KM and YT acknowledge partial support from the National Aeronautics and Space Administration (NASA) through NuSTAR Cycle-5 GO program (NNH18ZDA001N-NUSTAR).

VERITAS research is supported by grants from the U.S. Department of Energy Office of Science, the U.S. National Science Foundation and the Smithsonian Institution, by NSERC in Canada, and by the Helmholtz Association in Germany. This research used resources provided by the Open Science Grid, which is supported by the National Science Foundation and the U.S. Department of Energy's Office of Science, and resources of the National Energy Research Scientific Computing Center (NERSC), a U.S. Department of Energy Office of Science User Facility operated under Contract No. DE-AC02-05CH11231. We acknowledge the excellent work of the technical support staff at the Fred Lawrence Whipple Observatory and at the collaborating institutions in the construction and operation of the instrument.

The MDM Observatory is operated by Dartmouth College, Columbia University, The Ohio State University, Ohio University, and the University of Michigan.  We thank Justin Rupert and John Thorstensen for obtaining the optical spectra used in this paper. This research was supported by Basic Science Research Program through the National Research Foundation of Korea (NRF) funded by the Ministry of Science, ICT \& Future Planning (NRF-2017R1C1B2004566).

\facilities{MDM, \nustar, VERITAS}

\software{HEASoft (v6.28), \texttt{Stingray} \citep{Stingray-2016}, Eventdisplay (v4.83; \cite{Maier-2017}), NuSTARDAS (v1.8.0), \texttt{XSPEC} (v12.11.0; \cite{Arnaud-1996}), Naima \citep{Zabalza-2015}}

\end{document}